\documentclass[onecolumn,floats,showpacs,prd,amssymb,floatfix,nofootinbib,balancelastpage,superscriptaddress,amsmath,showkeys]{revtex4}

\usepackage{epsfig}
\usepackage{latexsym}
\usepackage{graphics}
\usepackage{graphicx}
\usepackage{amssymb}
\usepackage{epstopdf}
\usepackage{subfigure}
\newcommand{\ds}{{\sffamily DarkSUSY}}

\def\gtsima{$\; \buildrel > \over \sim \;$}
\def\ltsima{$\; \buildrel < \over \sim \;$}
\def\gsim{\lower.5ex\hbox{\gtsima}}
\def\lsim{\lower.5ex\hbox{\ltsima}}

\begin{document}

\title{Accurate estimate of the relic density and the kinetic decoupling in non-thermal dark matter models}

\author{Giorgio Arcadi}
\email{arcadi@sissa.it} 

\author{Piero Ullio}
\email{ullio@sissa.it}

\affiliation{SISSA,
              Via Bonomea 265, I-34136 Trieste, Italy and\\
              Istituto Nazionale di Fisica Nucleare,
              Sezione di Trieste, I-34136 Trieste, Italy}

%------------------------------------------------------------------------------

\begin{abstract}

Non-thermal dark matter generation is an appealing alternative to the standard paradigm of thermal WIMP dark matter. We reconsider 
non-thermal production mechanisms in a systematic way, and develop a numerical code for accurate computations of the dark matter relic 
density. We discuss in particular scenarios with long-lived massive states decaying into dark matter particles, appearing naturally in several 
beyond the standard model theories, such as supergravity and superstring frameworks. Since non-thermal production favors dark matter 
candidates with large pair annihilation rates, we analyze the possible connection with the anomalies detected in the lepton cosmic-ray flux 
by Pamela and Fermi. Concentrating on supersymmetric models, we consider the effect of these non-standard cosmologies in selecting 
a preferred mass scale for the lightest supersymmetric particle as dark matter candidate, and the consequent impact on the interpretation of new 
physics discovered or excluded at the LHC. Finally, we examine a rather predictive model, the G2-MSSM, investigating some of the standard
assumptions usually implemented in the solution of the Boltzmann equation for the dark matter component, including coannihilations. We 
question the hypothesis that kinetic equilibrium holds along the whole phase of dark matter generation, and the validity of the factorization 
usually implemented to rewrite the system of coupled Boltzmann equation for each coannihilating species as a single equation for the sum 
of all the number densities. As a byproduct we develop here a formalism to compute the kinetic decoupling temperature in case of
coannihilating particles, which can be applied also to other particle physics frameworks, and also to standard thermal relics
within a standard cosmology.

\end{abstract}



%------------------------------------------------------------------------------

\maketitle

\section{Introduction}

The identification of the dark matter (DM) component of the Universe is one of the most pressing issues in Science today.
Cosmological and astrophysical observations give compelling evidence for DM on a very wide range of scales; on the
other hand, from a particle physics perspective, these observations have not provided any clear indications on the relevant 
properties of DM particles, such as their mass and the interaction strength with ordinary matter (for a recent review on the DM 
problem and its particle physics implications, see, e.g.,~\cite{DMbook}). 
Viable DM candidates proposed in the literature include particles with a mass (close to) the Planck scale that are only 
gravitationally interacting, see e.g.~\cite{Chung:1998ua}, as well as ultra-light scalar particles possibly forming a condensate, 
see e.g.~\cite{Hu:2000ke}. The picture becomes more constrained once a mechanism to generate the DM term is considered. 
One of the most popular scenarios relies on a very general and elegant argument: stable massive species have an early-Universe 
thermal relic abundance scaling with the inverse of their pair annihilation rate, matching the cosmologically measured 
DM density when the annihilation cross section is about $3\cdot 10^{-26}\;{\rm cm}^3\,{\rm s}^{-1}$, a natural value for weak-force 
type couplings. This is the well-celebrated WIMP (weakly-interacting massive particle) miracle, allowing
to embed a DM candidate in most of the proposed extensions to the standard model (SM) of particle physics,
such as with the lightest neutralino in R-parity conserving supersymmetric (SUSY) extensions to the SM, or a heavy photon
in T-parity conserving versions of Little Higgs models~\cite{Birkedal:2006fz}. One aspect which is particularly appealing 
is the fact that, in most of these examples, the existence of such states and of the symmetry enforcing their stability are
not  properties  introduced ad-hoc to address the DM problem, but rather a byproduct of other features in the theory.  

Although extremely successful and attractive, the WIMP scenario faces also a few shortcomings. One the most severe is the fact  that
the idea of thermal generation of DM is based on the extrapolation of the properties of Universe from the earliest epoch at which the 
standard model for cosmology is well-tested, the onset of the synthesis of light elements (Big Bang Nucleosynthesis, BBN)
at a temperature $T_{BBN}$ of about 1~MeV, to the much earlier epoch of WIMP thermal freeze-out, at the temperature $T_{t.f.o.}$ of 
about one twentieth of the WIMP mass. The WIMP miracle relies on three main assumptions: {\sl i)} at freeze-out the Universe is 
in a radiation dominated phase with effective number of relativistic degrees of freedom as inferred from the SM particle spectrum; 
{\sl ii)}  there is no entropy injection intervening between $T_{t.f.o.}$ and $T_{BBN}$; {\sl iii)} there is no extra source of DM particles
on top of the thermal component. There are particle physics models in which all these three hypothesis are 
actually strongly violated, such as in any theory containing heavy states that are very weakly (e.g., gravitationally) coupled to ordinary 
matter, such as the gravitino or moduli fields in SUSY setups. These states do not thermalize in the early Universe, they may
dominate the Universe energy density, they are long-lived and potentially a copious source of entropy and DM particles at
decay. The so-called cosmological gravitino or moduli problem refers to the very severe observational limits one encounters
when these phenomena intervene during or after the BBN; on the other hand they can be perfectly consistent with available data
if the lifetime of these fields is shorter than the age of the Universe at the onset of BBN, about 1~s, or, equivalently, if the Universe 
is reheated to a temperature $T_{\rm RH}$ larger than $T_{BBN}$, where the reheating temperature $T_{\rm RH}$ is defined as the temperature
at which the Universe starts evolving in according to a radiation dominated phase after the field decays.

The prediction for the relic density of DM in case of long-lived heavy fields is generally very model dependent; there are however 
a few definite pictures. One of the most appealing is, e.g., the one pointed in Ref.~\cite{Moroi:1999zb} for SUSY DM in the 
anomaly-mediated SUSY breaking framework: it foresees the existence of a heavy modulus driving the Universe into a matter 
dominated phase and then decaying with a very large entropy production. The entropy injection reheats 
the Universe to a $T_{\rm RH}$ in the range between a few MeV and about 100~ MeV, and dilutes thermal relics (and gravitinos); 
in case the decay branching ratio into DM particles is unsuppressed, a very large number of DM particles is produced as well
but gets instantaneously reduced to level at which their pair annihilation rate matches the value of the Universe expansion rate at 
$T_{\rm RH}$. As a rule of thumb, the relic density of DM in this scenario turns out to be approximately equal to the result
for thermal relic WIMPs scaled up by the ratio $T_{t.f.o.}/T_{\rm RH}$: to be in agreement with the cosmological measurements of the DM 
density, the DM pair annihilation cross section needs also to increase accordingly (this solution to the moduli problem and the new match 
in the annihilation cross section appear naturally in some particle physics frameworks and is sometimes referred to as a 'non-thermal WIMP 
miracle'~\cite{Acharya:2009zt}; a more detailed discussion of the framework and a list of relevant references is given below).

The interest in DM particles with large pair annihilation cross section has been recently triggered by the fact that cross sections larger 
than the standard face-value for WIMP DM are needed to provide a DM positron source accounting for the rise in the positron fraction 
in the local cosmic-rays measured by the PAMELA detector~\cite{Adriani:2008zr}. The picture of non-thermal generation of DM has 
however a much broader phenomenological impact, e.g., shifting significantly the mass scale for which a DM particle embedded in
a SM extension is cosmologically relevant or excluded, with a direct impact as well on the interpretation of new physics discovered or 
excluded at the LHC.

In the paper we reexamine the issue of non-thermal generation of DM implementing a full numerical solution of the relevant set of 
equations, including the equation of motion for the heavy fields and the system of coupled Boltzmann equation, and avoiding to 
introduce approximations such as instantaneous reheating and production of DM particles, or others. 
The code is interfaced to an appropriately modified version of the 
public available \ds\ package~\cite{Gondolo:2004sc} and allows, for any definite particle physics scenario, a very accurate computation
of the relic abundance of the DM particle. Examples of its applications are given in a sample of progressively refined particle 
physics scenarios: we first consider a toy-model in which DM particles are schematically defined  through the values of the mass and
a temperature-independent pair annihilation cross section. We examine then the impact of heavy long-lived fields on the phenomenology
of the most widely studied case for WIMP DM, i.e. the case of neutralino dark matter in the minimal supersymmetric extension of the
standard model (MSSM), focussing in particular on the framework usually dubbed "Split Supersymmetry"~\cite{ArkaniHamed:2004fb,Giudice:2004tc},
in which, since the sfermion sector is not playing any relevant role, the parameter space is sharply reduced compared to a general MSSM,
but it is still general enough for our purpose. Finally we examine in detail the G2-MSSM scenario~\cite{Acharya:2007rc}, a  particular class 
of theories with rather precise predictions for the spectrum of low energy SUSY particles as well as of the gravitino and the moduli fields; 
in such a definite framework it is interesting to test the validity of some of the approximations that are usually given for granted in the
solution of the Boltzmann equation for DM: In this model the lightest neutralino is the lightest SUSY particle but it also nearly degenerate in 
mass with a chargino; coannihilations, namely the interplay between the two in the early Universe, are usually treated by writing a single
Boltzmann equation for the sum of the two species, however for very low reheating temperatures this approach may not be valid. 
We solve the full system of coupled Boltzmann equation and address also the issue of energy losses for relativistic neutralinos
and charginos injected by moduli decays. As a byproduct and final step, we develop here for the first time the formalism to compute the 
DM kinetic decoupling  temperature for a system undergoing a low-temperature reheating and for which kinetic equilibrium is maintained 
in a chain of coupled processes rather than by the elastic scattering of a single DM particle on thermal bath particles. This is an important 
result since thermal kinetic decoupling, namely when DM scattering goes out of equilibrium (as opposed to the thermal chemical decoupling 
mentioned above which refer to the departure from equilibrium of the pair annihilation processes) determines the small-scale 
cutoff in the spectrum of matter density fluctuations, see, e.g.,\cite{Profumo:2006bv,Bringmann:2006mu}.

The paper is organized as follows:  In Section 2 we review the particle physics scenario and illustrate the approach we follow 
to trace the evolution of the long-lived heavy fields and compute the relic density of DM particles. In Section 3 we give a first example of
this procedure introducing a simple toy-model; for this model we cross check various approximate scalings for the DM relic 
density discussed in the literature, and also derive, under different configurations,  what range of reheating temperature would be needed to provide 
an explanation for the very large annihilation cross section which may be inferred from the PAMELA positron excess. 
In Section 4, still within a schematic treatment of the decaying heavy fields, we consider the case of split SUSY in the MSSM and 
discuss the impact of low reheating temperatures on the cosmologically relevant portion of its parameter space. In Section 5, we
focus on the G2-MSSM, with the numerical treatment of the sequence of steps induced by the set of heavy long-lived fields present
in the model; the treatment of this case is refined in Section 6, where we also compute the DM kinetic decoupling temperature
for this model.

\section{The general framework for non-thermal dark matter production}
\label{sec:general}

We consider a particle physics framework embedding three sets of beyond-SM states: Let $\{{\chi_a}\}$ be a set of particles sharing 
a conserved quantum number and having sizable couplings with SM particles;  the first property ensures that the lightest of
them, say  $\chi_0$, is stable, while the second guarantees thermalization at sufficiently large temperatures. We also require 
that $\chi_0$ has zero electric and color charges, so that it can play the role of DM candidate. The fields $\{X_i\}$ in the second set,
which we will refer to as cosmological moduli,
are instead heavy states which are very weakly interacting, out of thermal equilibrium in the early Universe and long-lived;
we assume they can condensate, potentially dominate the Universe energy density at an intermediate stage in its evolution, and later 
decay producing both SM particles, with a sharp increase in the entropy density, and $\chi_a$ fields. Finally the framework may contain 
also additional long-lived states, say $\psi_i$, out of thermal equilibrium but with a subdominant contribution to the energy density, 
possibly sharing the quantum number protecting the stability of $\chi_0$; these particles may also be 
produced in the decay of the $X_i$ states.

A typical scenario of this kind is provided by SUSY extensions of the SM. We will consider in particular cases in which the states
$\{\chi_a\}$ are the superpartners of SM particles within the MSSM, and R-parity is the symmetry protecting the stability of the lightest
SUSY particle (LSP), which we will assume to be the lightest neutralino. For what regards the moduli $\{X_i\}$, there are several 
possibilities. First of all, it is quite common, in SUSY theories, to find field configurations for which the scalar potential is flat; these 
configurations are referred as 'flat directions' and can be described by a chiral superfield. For our purposes, only the scalar component 
of these multiplets is relevant; we will refer to this part with the term modulus. SUSY breaking can lift the flat directions inducing a 
mass term for the moduli in the scalar potential. Further candidates for long-lived states arise in supergravity theories. The gravitino 
is not in thermal equilibrium in the early Universe, however, plays a different role compared to moduli fields; we will restrict to the 
case when gravitinos  are heavy and not the LSP (otherwise the phenomenology would be very different from the one discussed 
in this paper), falling in the category of the $\psi_i$ fields introduced above. Another possibility is the Polonyi 
field~\cite{Coughlan:1983ci,Dine:1983cu,Ellis:1986zt}  which is introduced in many SUSY breaking schemes. 
Finally supergravity can be seen as a low energy limit of string theory, in which scalar fields can appear in the compactification of 
extra dimensions.

The impact on cosmology of the moduli can easily sketched under a few simplifying assumptions. Consider for simplicity the
cases of a single modulus $X$ which decays when it  dominates the Universe energy density; focussing on Planck suppressed interactions, 
its decay width can be written in the form:
\begin{equation}
  \Gamma_X = D_X \frac{m_X^3}{M_{Pl}^2}\,,
  \label{eq:dw}
\end{equation}
with $D_X$ some coefficient depending on the specific model, $m_X$ the field mass and $M_{Pl}$ the reduced Planck mass.
Assuming instantaneous conversion of the energy density 
into radiation, one usually defines the reheating temperature $T_{\rm RH}$ through the expression:
\begin{equation} 
  \label{eq:trh}
  \Gamma_X \equiv  \sqrt{\frac{\pi^2  g_{\rm eff}(T_{\rm RH})}{90}} \frac{T_{\rm RH}^2}{M_{Pl}}\,,
\end{equation}
where $g_{\rm eff}(T_{\rm RH})$ is the effective number of relativistic degrees of freedom at $T_{\rm RH}$. Inverting this 
expression, one finds approximately that the onset of the standard radiation dominated phase happens at the temperature:
\begin{equation} 
  \label{eq:trh2} 
  \frac{T_{\rm RH}}{1\,{\rm MeV}} \simeq 0.62  D_X^{1/2} \left[\frac{10.75}{g_{\rm eff}(T_{\rm RH})}\right]^{1/4} \left(\frac{m_X}{10\,{\rm TeV}}\right)^{3/2}\,.
\end{equation}
To avoid spoiling predictions of the standard BBN, one needs to require $T_{\rm RH} \gsim 4$~MeV~\cite{Kawasaki:2000en}, which, for $D_X$ of order
one, translates into a lower limit on the mass of the cosmological modulus of about 30~TeV.

At this level of approximation the evolution of the system would be fully specified by the $X$ decay width (or $T_{\rm RH}$) and the amount of energy 
density converted into dark matter particles (which, in the treatment above, was implicitly assumed to be tiny compared to amount going into radiation). 
Having instead in mind to be able to treat a system in which, for the full set of $X_i$ fields, spectra, lifetimes and branching ratios in the decay are calculable
in the given particle scenario, we will not refer to instantaneous reheating, but rather follow more explicitly the evolution of the moduli. In principle this could be 
done by studying a full set of coupled equations of motion, having specified the potentials for each field~\cite{Dine:1995kz,Acharya:2008zi}.
 The result would be that, at early times, each field stays frozen in a time dependent minimum; when ${t}^{-1}$ becomes of the order of the 
$X_i$ mass, the equation of motion takes the form of the one for a damped harmonic oscillator. The oscillations satisfy a pressureless equation of state and 
hence the scalar field behaves like a condensate evolving as a matter fluid; provided that enough energy is initially stored in $X_i$, the Universe enters 
a phase of matter domination lasting until the field decays, with the transition that needs to be treated as a continuos process. In practice, even for models for 
which the physics related to the $X_i$ fields is given in some detail, it is difficult to describe potentials and their temperature evolutions beyond the toy model 
level; for our purposes it will be sufficient to follow the evolution of the system starting from the phase of coherent oscillations. Each state $X_{i}$ is then traced 
through an equation for its energy density:
\begin{equation}
  \label{eq:mod}
  \frac{d\rho_{X_{i}}}{dt} + 3H\rho_{X_{i}} = -\Gamma_{X_i}\rho_{X_i}\,,
\end{equation}
and in case of several moduli present at the same time, the single equations are included in the system at the time $t = 3/2 H =  m_{X_i}$, assuming the energy 
density stored in the field at this time is equal to $(1/2)m_{X_i}^2 M_{Pl}^2$\cite{Giudice:2000ex,Dine:1995kz,Acharya:2008zi}; 
we will comment later on the fact that the final density of dark matter
is not sensitive to these assumptions.

The decay of the $X_i$ particles produces SM particles, ${\chi}_a$ states cascading to the DM particle, and, eventually, the long-lived $\psi_i$ fields, in turn 
decaying into radiation and, possibly, DM particles. From the first principle of thermodynamics, one can write an equation 
for the total energy density and pressure associated to SM, $\chi_{a}$ and ${\psi}_{j}$ states, respectively, $\rho$ and $p$,
in an implicit form:
\begin{equation}
  \label{eq:firstlaw}
  \frac{d \rho}{dt}+3H(\rho+p)=\sum_i \Gamma_{X_i} \rho_{X_i}\,.
\end{equation}
This equation is treatable once separating $\rho$ and $p$ in components. 
Starting with the ${\psi}_{j}$ particles, one can safely assume that they are produced in given number at decays, get diluted and redshifted by the 
Universe expansion without interacting with other species and decay themselves (an eventual term  associated to the production via inelastic scattering off SM 
or ${\chi}_a$ particles is not introduced since such term becomes relevant only at large temperatures, while we will only consider here the case of moderate to low 
$T_{\rm RH}$; also, we are not considering the possibility of a  ${\psi}_{j}$ particle decaying into a lighter  ${\psi}_{k}$ state, since we will not encounter a case 
of this kind in explicit models and it would just complicate the notation). The Boltzmann equation for  the ${\psi}_{j}$ number density is:
\begin{equation}
  \label{eq:psi}
  \frac{dn_{\psi_j}}{dt}+3H n_{\psi_j}=\sum_i \frac{B_{\psi_j,X_i}}{m_{X_i}} \Gamma_{X_i} \rho_{X_i} - \Gamma_{\psi_j} n_{\psi_j}\,,
\end{equation}
where $B_{\psi_j,X_i}$ is the mean number of particles $\psi_j$ generated in the decay of the field $X_i$, i.e. it is the product of the branching ratio of decay into ${\psi}_{j}$
times the mean multiplicity. 

To trace the number density of the $\chi_a$ states, especially when two or more of these are nearly degenerate in mass (coannihilating particles), 
one should refer to a system of coupled Boltzmann equations describing: the source from the
decay of the $X_i$ and $\psi_j$ fields; their changes in number density due to pair production from and annihilation into SM particles; 
the energy exchanges with SM thermal bath particles through elastic scatterings processes; the redistribution in the relative 
number density by  inelastic scattering of a given $\chi_a$ into a different $\chi_b$ state; the decays of  $\chi_c$ into lighter $\chi_d$ particles
and and of these to the lightest stable species. This is usually not done since it is a system of coupled stiff equations one needs to solve numerically; 
moreover it is usually not necessary to do it, since one is interested only in the number density of the lightest state after all heavy states have 
decayed into the stable one. Rather than tracing the number density $n_{\chi_a}$ of the individual state $\chi_a$, one usually solves a single equation 
written for the sum of all the number densities, $n_\chi = \sum_a n_{\chi_a}$, i.e.~\cite{Griest:1990kh,Edsjo:1997bg}:
\begin{equation}
   \label{eq:dm}
   \frac{dn_\chi}{dt} + 3\,H\,n_\chi = - \langle\sigma_{\rm{eff}} v\rangle \left[n_\chi^2 - (n_\chi^{eq})^2 \right] + \sum_i \frac{B_{X_i}}{m_{X_i}} 
   \Gamma_{X_i} \rho_{X_i} + \sum_j B_{\psi_j} \Gamma_{\psi_j} n_{\psi_j}
\end{equation}
where $B_{X_i} \equiv \sum_a B_{\chi_a,X_i}$ and $B_{\psi_j} \equiv \sum_a B_{\chi_a,\psi_j}$ have been defined in analogy to $B_{\psi_j,X_i}$. In this equation
$n_\chi^{eq}$ stands for the sum of thermal equilibrium number densities, and the term proportional its square accounts for the production of particles $\chi_a$ in
pair annihilations of SM thermal bath particles, while the effective thermally averaged annihilation cross section: 
\begin{equation}
  \label{eq:sum}
  \langle\sigma_{\rm{eff}} v\rangle = \sum_{a,b} \langle\sigma_{ab} v_{ab}\rangle \frac{n_{\chi_a}^{eq}}{n_\chi^{eq}} \frac{n_{\chi_b}^{eq}}{n_\chi^{eq}}\, 
\end{equation}
is written as a weighted sum over the thermally averaged annihilation cross section of any $\chi_a$-$\chi_b$ pair into SM particles; 
the processes giving a sizable contribution to this sum are only those for which the mass splitting between a state $\chi_a$ 
and the lightest state $\chi_0$ are comparable to the thermal bath temperature $T$. 
There are two main assumptions which allow to implement Eq.~(\ref{eq:dm}) to trace $n_\chi$. The first is kinetic equilibrium for each species $\chi_a$,
namely that the scattering processes on thermal bath particles are efficient and make the phase space densities for each particle trace the spectral shape of the 
corresponding thermal equilibrium phase space density, namely  $f_a(k_a,t)= C(t) \cdot f_a^{eq}(k_a,t)$ (with the coefficient $C$ depending on time but not on 
momentum). Within this assumption, we treat as instantaneous the energy depletion from the relativistic regime when particles are injected from moduli or  ${\psi}_{j}$ 
decays to the non-relativistic velocities in the low reheating temperature background plasma; it also allows to factorize the number density 
out of each thermally averaged annihilation cross section (which is defined in terms of  thermal equilibrium phase space densities). To implement the factorization
of the individual terms in the sum of Eq.~(\ref{eq:sum}) one needs also to assume that $n_{\chi_a}/n_\chi \simeq n_{\chi_a}^{eq}/n_\chi^{eq}$, a quantity which, 
in the Maxwell-Boltzmann approximation for the equilibrium phase space densities, as appropriate for non-relativistic particles, is proportional to the number of 
internal degrees of freedom $g_{\chi_a}$ and is exponentially suppressed with the ratio between mass splitting and temperature; this approximation is strictly valid 
only in case inelastic scatterings of $\chi_a$ particles are efficient over the whole time interval in which the pair annihilation term is relevant. 
Within the standard computation of the thermal relic density for WIMPs, the two assumption are in general well justified, since the kinetic decoupling and the
decoupling of inelastic scatterings usually take place at a much lower temperature than chemical decoupling; while assuming that  Eq.~(\ref{eq:dm}) is valid
in the next Sections, in Section~6 we study this issue in more details and, considering a specific particle physics scenario, address the problem of kinetic decoupling 
in models with non-thermal generation of DM particles.

We keep track of the SM states only through their contribution to the radiation energy density and pressure which, using Eq.~(\ref{eq:firstlaw}) and 
subtracting the contribution from $\psi_j$ and $\chi_a$ fields, obey the equation:
\begin{eqnarray}
  \label{eq:rad}
  \frac{d\rho_R}{dt}+3H (\rho_R+p_R) & \simeq & \sum_i \left(1 - \frac{\sum_j B_{\psi_j,X_i}\,\langle E_{\psi_j,X_i}\rangle + B_{X_i}\,m_\chi}{m_{X_i}} \right) 
  \Gamma_{X_i} \rho_{X_i}  +  \sum_j \left( \langle E_{\psi_j} \rangle - m_\chi B_{\psi_j} \right) \Gamma_{\psi_j} n_{\psi_j}  \nonumber\\
  &&  + m_\chi  \langle\sigma_{\rm{eff}} v\rangle \left[n_\chi^2 - (n_\chi^{eq})^2 \right]\;.
\end{eqnarray}
In this equation $\langle E_{\psi_j,X_i}\rangle$ is the mean energy of the particle ${\psi}_{j}$ at injection from the decay of the modulus $X_{i}$:
\begin{equation}
  \langle E_{\psi_j,X_i}\rangle \equiv \int dE^\prime \, \frac{d{\mathcal N}_{\psi_j,X_i}}{dE^\prime}\,E^\prime
\end{equation}
with $d{\mathcal N}_{\psi_j,X_i}/dE^\prime$ the energy spectrum from the decay normalized to 1; $\langle E_{\psi_j} \rangle$ is instead the mean energy
for ${\psi}_{j}$ particles:
\begin{eqnarray}
  \langle E_{\psi_j} \rangle(t) \, n_{\psi_j}(t) &\equiv &\int_0^t dt^\prime \sum_i \frac{B_{\psi_j,X_i}}{m_{X_i}} \Gamma_{X_i} \rho_{X_i}(t^\prime) 
  \int dE^\prime \, \frac{d{\mathcal N}_{\psi_j,X_i}}{dE^\prime}\,\left[m_{\psi_j}^2+\frac{a^2(t^\prime)}{a^2(t)}({E^\prime}^2-m_{\psi_j}^2) \right]^{1/2} \cdot \\ \nonumber
  && \cdot \frac{a^3(t^\prime)}{a^3(t)} {\rm exp}\left[- \Gamma_{\psi_j} (t-t^\prime)\right]\;.
\end{eqnarray}
Finally, in Eq.~(\ref{eq:rad}) we have assumed that the mean energy of the $\chi_a$ states is equal to the mass $m_\chi$ of the lightest state $\chi_0$, neglecting, 
at this level, thermal corrections and mass splittings between the coannihilating states, as well as the pressure term associated to $\chi_a$. 

Eqs.~(\ref{eq:mod}), (\ref{eq:psi}), (\ref{eq:dm}) and (\ref{eq:rad}) define a system of coupled equations, closed by Friedmann equation giving $H$. In its numerical 
solution, it is more convenient to use as independent variable, rather than the time $t$, the rescaled scale factor $A \equiv a/a_I$, with $a_I$ an arbitrary parameter 
with dimension of the inverse of an energy. Following~\cite{Giudice:2000ex}, we will use as dependent variables the dimensionless quantities:
\begin{equation}
  \xi_{X_i} \equiv \frac{\rho_{X_i} a^3}{\Lambda}\,, 
  \quad \quad  N_{\psi_j} \equiv n_{\psi_j} a^3\, \quad \quad {\rm and} \quad \quad  N_\chi \equiv n_\chi a^3\,,
\end{equation}
with $\Lambda$ an arbitrary energy scale, plus the temperature $T$, expressing $\rho_R$ and $p_R$ in terms of the entropy density through the standard definitions:
\begin{equation}
  s(T)=\frac{\rho_R(T)+p_R(T)}{T} \equiv \frac{2\pi^2}{45} h_{\rm eff}(T) T^3 \quad \quad {\rm and} \quad \quad 
  \rho_{R}(T) \equiv \frac{\pi^2}{30} g_{\rm eff}(T) T^4 = \frac{3}{4} \frac{g_{\rm eff}(T)}{h_{\rm eff}(T)} T s(T) \,,
\end{equation}
with $g_{\rm eff}$ and $h_{\rm eff}$ the effective number of relativistic degrees of freedom. The values of ${a}_{I}$ and $\Lambda$ are chosen in order to guarantee 
the best numerical stability to the solution, a sample guess being, respectively, $T_{\rm RH}^{-1}$ and $T_{\rm RH}$, with the approximate reheating scale as given 
through Eq.~(\ref{eq:trh}). After this change of variables the system becomes:
\begin{eqnarray}
  \label{eq:system}
  \frac{d{\xi}_{X_i}}{dA} & = &-\frac{A^{1/2} a_I^{3/2}}{\mathcal H} \Gamma_{X_i} \xi_{X_i} \\  \nonumber
  \frac{dN_{\psi_j}}{dA} & = & \frac{A^{1/2} a_I^{3/2}}{\mathcal H} 
         \left(  \Lambda \sum_i \frac{B_{\psi_j,X_i}}{m_{X_i}} \Gamma_{X_i}  \xi_{X_i}- \Gamma_{\psi_j} N_{\psi_j}\right) \\ \nonumber
  \frac{dN_\chi}{dA} & = &  - \frac{\langle\sigma_{\rm{eff}} v\rangle}{A^{5/2} a_I^{3/2} {\mathcal H}} \left[N_\chi^2 - (N_\chi^{eq})^2 \right]
         + \frac{A^{1/2} a_I^{3/2}}{\mathcal H} \left( \Lambda \sum_i \frac{B_{X_i}}{m_{X_i}} \Gamma_{X_i} \xi_{X_i} 
         + \sum_j B_{\psi_j} \Gamma_{\psi_j} N_{\psi_j} \right) \\ \nonumber
  \frac{dT}{dA} & = & {\left(1+\frac{T}{4 g_{\rm eff}}\frac{dg_{\rm eff}}{dT}\right)}^{-1}
  \left\{-\frac{h_{\rm eff}}{g_{\rm eff}} \frac{T}{A} + \frac{h_{\rm eff}}{3 g_{\rm eff} s(T)} \frac{1}{A^{5/2} a_I^{3/2} {\mathcal H}} \right. 
  \left[  \sum_j \left( \langle E_{\psi_j} \rangle - m_\chi B_{\psi_j} \right) \Gamma_{\psi_j} N_{\psi_j} \right. \\ \nonumber
  && \left. \left.
  +  \Lambda \sum_i \left(1 - \frac{\sum_j B_{\psi_j,X_i}\,\langle E_{\psi_j,X_i}\rangle + B_{X_i}\,m_\chi}{m_{X_i}} \right) 
  \Gamma_{X_i} \xi_{X_i}  + \frac{m_\chi  \langle\sigma_{\rm{eff}} v\rangle}{A^3 a_I^3} \left[N_\chi^2 - (N_\chi^{eq})^2 \right] \right] \right\}
\end{eqnarray}
where ${\mathcal H}$ is defined from the Universe expansion rate, as:
\begin{equation}
  {\mathcal H} \equiv (a_I A)^{3/2} H = \left(\frac{\Lambda \sum_i \xi_{X_i}  + \rho_R(T) A^3 a_I^3 + m_\chi N_\chi + \sum_j  \langle E_{\psi_j} \rangle
  N_{{\psi}_{j}}}{3M^2_{PL}}\right)^{1/2}\,.
\end{equation}
The relic density of dark matter can be evaluated by evolving these equations from an initial time, which we assume to be the time when the heaviest modulus starts
its coherent oscillations, up to the stage when the DM comoving number density becomes constant. 

\section{Non-thermal DM production in a toy model and relevance for Pamela}

We discuss first a minimal framework with a single cosmological modulus $X$ decaying into the DM particle $\chi$. Rather than detailing a specific 
particle physics scenario, in this first example we define $\chi$ only through its mass and pair annihilation rate into SM particles, whose thermal average
is assumed not to depend on temperature, as appropriate for S-wave annihilations. We also avoid dealing with eventual other states charged under the 
quantum number protecting the stability of $\chi$, assuming that they have a sizable mass splitting with respect to $\chi$, and hence have very short 
lifetimes and do not enter in the Boltzmann equation for $\chi$. 
Under these hypotheses the system of coupled equations reduces to three equations only: the first for the decaying modulus, the second for number density
of particle $\chi$, sourced from the decay and depleted by pair annihilations, and the last for the temperature.

In this simplified picture, the main trends in the non-thermal DM production can be illustrated even at the level of approximate analytical formulae; we briefly 
summarize here some of these features, as we will recover them in the numerical solution of this model as well as in the more involved scenarios we will consider later
(for a more detailed discussion, see, e.g., \cite{Gelmini:2006pw}). First of all, if the modulus decay induces a large increase in the entropy density and this happens
at a later stage with respect to the chemical decoupling for $\chi$, the thermal relic density of $\chi$ is greatly diluted and can be neglected, with the only relevant
$\chi$ source being the particles produced in the decay itself. The entropy injection is a continuos process making the reheating phase last for an extended period during which one can show that the temperature evolves as ${T}\propto {a}^{-3/8}$ and the universe expansion rate as $H \propto T^4$~\cite{Giudice:2000ex,Gelmini:2006pw}.
A standard approximation is however to treat the decay of the field and the thermalization of the products as instantaneous processes, and define the reheating 
temperature $T_{\rm RH}$ according to Eq.~(\ref{eq:trh}); depending on whether at $T_{\rm RH}$ the dark matter pair annihilation rate $\Gamma = n_\chi  \langle \sigma v \rangle$ 
is larger or smaller than the expansion rate $H$, there are two distinct regimes determining the relic density for $\chi$~\cite{Moroi:1999zb,Gelmini:2006mr}. 
If $\Gamma$ is much larger than $H$, pair annihilations are very efficient and instantaneously decrease in the number density of $\chi$ to the 
critical density level corresponding to $\Gamma \simeq H$ when the annihilations stop; such critical density is then simply equal to:
\begin{equation}
  \label{critical}
  n_\chi^c \simeq  \frac{H}{\langle \sigma v \rangle} \,.
\end{equation}
As usually done, we normalize the number density to the entropy density introducing the quantity $Y_\chi=n_\chi/s$, since when annihilations become inefficient,  
if there are no further entropy injection phases, such ratio becomes constant and can be used to estimate the relic density for $\chi$:
\begin{equation}
  \Omega_\chi^{NT} = \frac{m_\chi s(T_0)}{\rho_c(T_0)} Y_\chi(T_0) =   \frac{m_\chi s(T_0)}{\rho_c(T_0)} Y_\chi(T_{\rm RH}) \propto \frac{m_\chi} {\langle \sigma v \rangle\, T_{\rm RH}}\,,
\end{equation}
where $\rho_c(T_0)$ and $s(T_0)$ refer to the Universe critical density and entropy density at present.The rule of thumb $\Gamma \simeq H$ is the same criterium implemented for an approximate estimate of the relic density in the standard thermal decoupling picture for WIMP dark matter, except that the reference temperature in this latter case is the thermal freeze-out temperature ${T}_{t.f.o.}\simeq m_\chi/20$. Following the same steps, one finds that the thermal relic density ${\Omega}_\chi^{T}$ scales with the inverse of ${T}_{t.f.o.}$, and hence that the relations of ${\Omega}_\chi^{NT}$  with ${\Omega}_\chi^{T}$ and the WIMP pair annihilation cross section are approximately given by:
\begin{equation}
  \label{eq:ntc}
  \Omega_\chi^{NT} h^2 \simeq \frac{{T}_{t.f.o.}}{T_{\rm RH}}\, \Omega_\chi^{T} h^2 \simeq \frac{m_\chi/20}{T_{\rm RH}} \cdot
  \frac{3\cdot 10^{-26}\;{\rm cm}^3\,{\rm s}^{-1}}{\langle \sigma v \rangle}\,. 
\end{equation}
A particle $\chi$ whose thermal relic density is small compared to the DM density because the annihilation rate is too large, may become a viable dark matter candidate for
an appropriate value of $T_{\rm RH}$. This simple rescaling holds whenever the particles $\chi$ are copiously produced in the modulus decay and if the pair annihilation rate is 
sufficiently large; in the following, we refer this scenario as 'reannihilation regime'. If instead $n_\chi(T_{\rm RH})$ is lower than $n_\chi^c(T_{\rm RH})$, the particles produced in the decay do not interact further and their number density per comoving volume is frozen, being:
\begin{equation}
  Y_\chi(T_{\rm RH}) = \frac{n_\chi(T_{\rm RH})}{s(T_{\rm RH})} \simeq \frac{B_X}{m_X}  \frac{\rho_X(T_{\rm RH})}{s(T_{\rm RH})} 
  \simeq \frac{3}{4} \frac{B_X}{m_X} T_{\rm RH}
\end{equation}
and hence giving a non-thermal relic density which is about (see also, e.g., \cite{Gelmini:2006pw}): 
\begin{equation}
  \label{eq:ntnc}
  \Omega_\chi^{NT} h^2 \simeq 0.2 \cdot 10^4 \, B_X \frac{10\,{\rm TeV}}{m_X} \frac{T_{\rm RH}}{1\,{\rm MeV}} \frac{m_\chi}{100\,{\rm GeV}}  
\end{equation}
Note that, in this case, the final dark matter density depends on the physics of moduli not only through its proportionality to the reheating temperature 
but also through the ratio between the average number of particles $\chi$ produced per decay and the modulus mass $B_X/m_X$. 
This non-thermal scaling applies to the cases in which either the pair annihilation rate is small or the average number of particles $\chi$ produced per 
decay $B_X$ is small.\footnote{Ref.~\protect{\cite{Gelmini:2006pw}} classifies two extra scenarios, already studied, e.g., in Ref.~\cite{Giudice:2000ex}, corresponding to
the case in which the main source of $\chi$ particles is pair production from SM background states; these applies essentially only in the limit of 
$B_X\rightarrow 0$ which we are not going to discuss, although the method outlined here would be suitable for them as well.}

We are now ready to discuss numerical results within this simplified scenario. As just outlined, the relevant parameters for the relic density calculation are the particle 
mass and pair annihilation cross section, as well as those setting the efficiency in producing dark matter particles and the energy density of the field at decay; regarding
the latter we will treat as free parameters $B_X$ and the mass of the modulus $m_X$, which in turns sets the decay width $\Gamma_X$ and hence the reheating energy 
(we start with the assumption of gravitational interactions in the decay, and comment shortly on how to interpret results in case of a more general
expressions for $\Gamma_X$). Since we are tracing the full evolution of the field $X$, we are not in the limit of instantaneous reheating and do not implement the
definition of reheating temperature as quoted in Eq.~(\ref{eq:trh}); the $T_{\rm RH}$ we refer to when illustrating results is 
extrapolated from the numerical solution, matching the ${T} \propto {a}^{-3/8}$ scaling obtained in the phase when the $X$ decays act as a large source 
of entropy to the ${T} \sim {a}^{-1}$ scaling in the subsequent radiation dominated regime (this prescription of matching asymptotic solutions is not totally 
rigorous since we should also take into account eventual changes in the number of relativistic degrees of freedom contributing to the entropy density; in 
practice, however, since the transition between the two regimes is always rather sharp, the $T_{\rm RH}$ found in this way is always very accurate in parametrizing the total entropy injection from the $X$ decay; note also that $T_{\rm RH}$ is not used in any step of the numerical computation).

\begin{figure}[t]
 \begin{minipage}[htb]{8cm}
   \includegraphics[width=8 cm, height= 6 cm, angle=360]{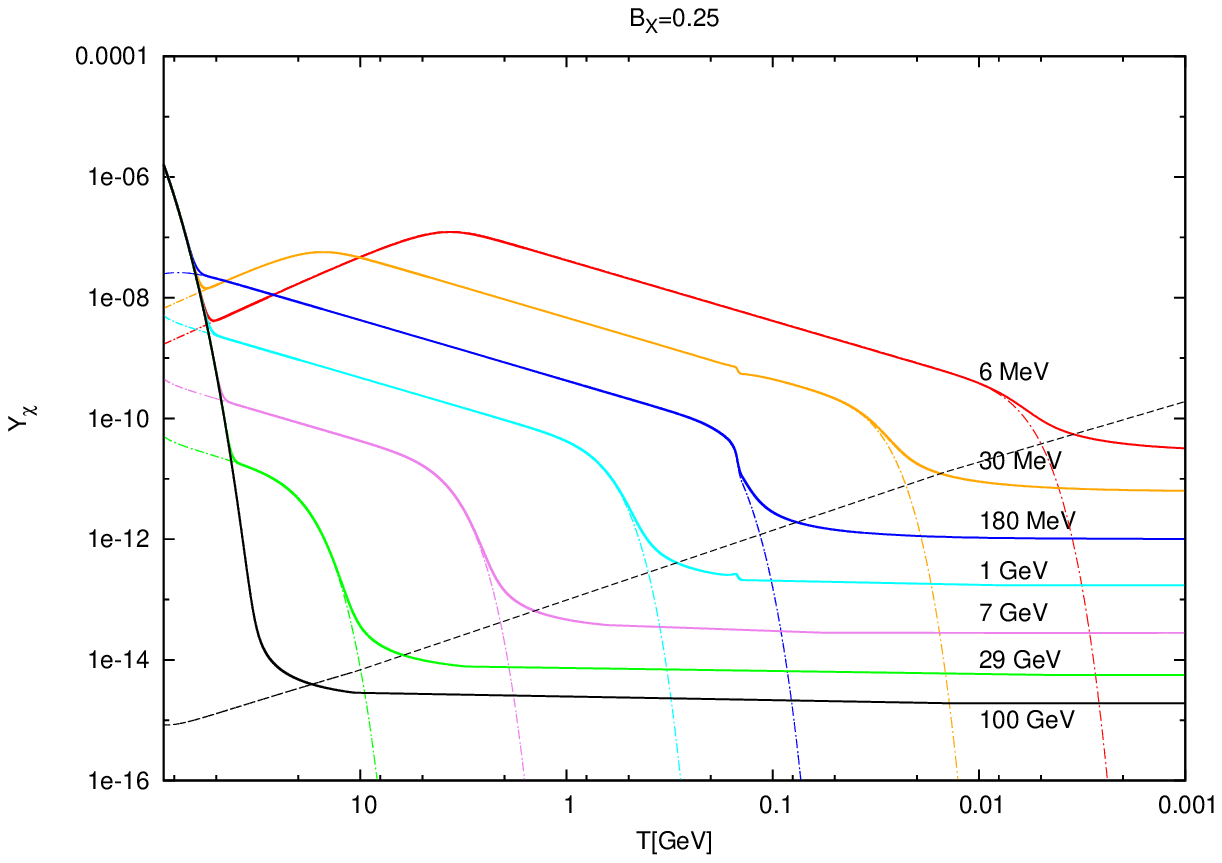}
 \end{minipage}
 \ \hspace{3mm} \
 \begin{minipage}[htb]{8cm}
   \includegraphics[width=8cm, height= 6 cm, angle=360]{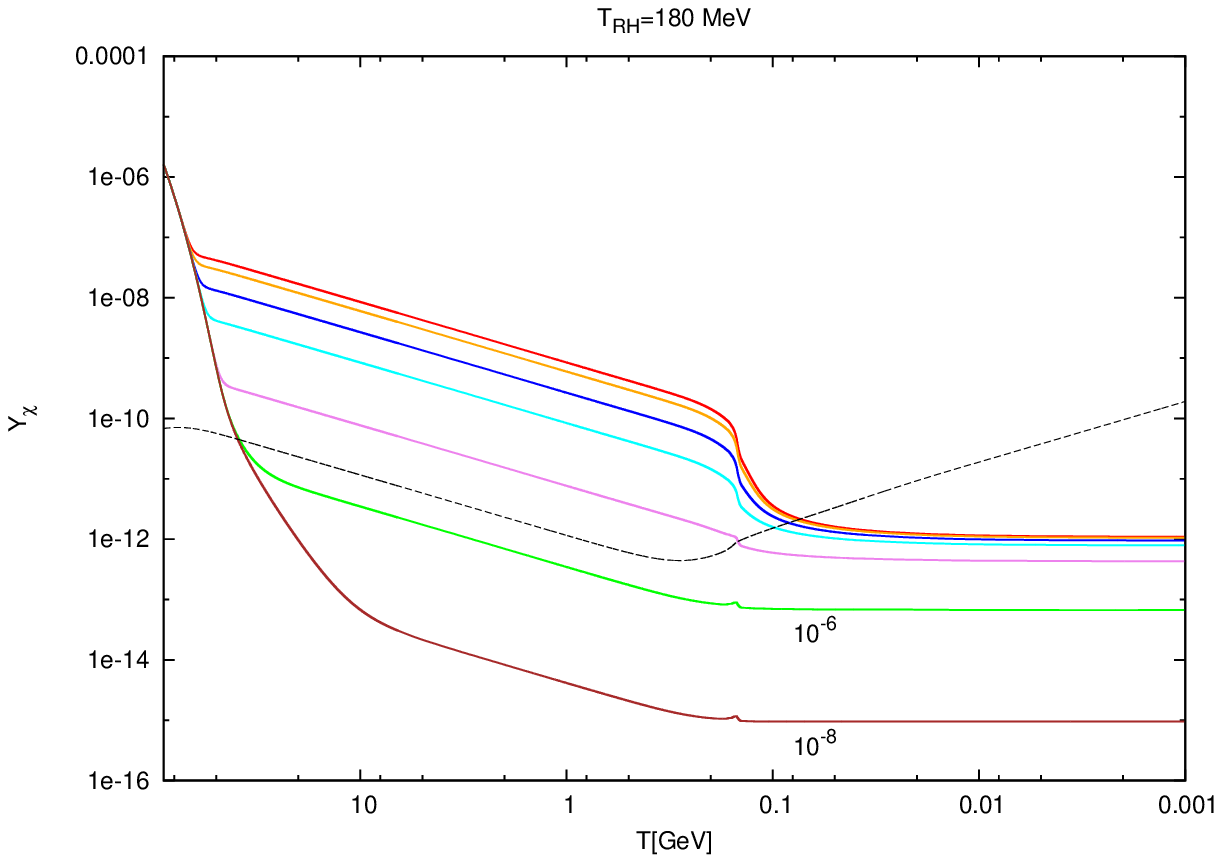}
 \end{minipage}
 \caption{The number density of the DM particles normalized the entropy density as a function of temperature, for few values of the reheating temperature 
  and $B_X$ fixed to 0.25 (left panel), and for a sample $T_{\rm RH}$ while varying $B_X$ (right panel). In the left panel we also plot with dashed-dotted lines the 
  quasi-equilibrium density $Y_\chi^{QSE}(T)$ for each $T_{\rm RH}$, and with a dashed line the critical density $Y_\chi^c(T)$ for the case $T_{\rm RH} = 100$~GeV
  in which reheating takes place before thermal freeze out. In the right panel the dashed line shows $Y_\chi^c(T)$ for $T_{\rm RH} = 180$~MeV. 
  Results refer to a DM particle with mass and pair annihilation cross section being, respectively, 1~TeV and $5 \cdot 10^{-24}$~cm$^3$~s$^{-1}$. }
\label{fig:fig1}
\end{figure} 

In Fig.~\ref{fig:fig1} we consider a sample DM particle with heavy mass, $m_\chi = 1$~TeV, and large pair annihilation cross section, 
$\langle \sigma v \rangle = 5 \cdot 10^{-24} {\mbox{cm}}^{3}{\mbox{s}}^{-1}$; the ratio $Y_\chi$ of the DM number density to the entropy density is
plotted as a function of $T$ (note that, since we want to compare directly $T_{\rm RH}$ with $T$, rather than showing $Y_\chi$ versus the inverse of 
temperature as usually done, we plot it versus $T$ and use a logarithmic scale which decreases from left to right). 
In the left panel we have fixed $B_X$ to a sample value representative of the case when the branching ratio of the decay into $\chi$ is 
unsuppressed, and vary $m_X$ to select a few values of the reheating temperature; in the right panel, vice versa, we fix $m_X$ and vary $B_X$. 
The system of equations is solved assuming the initial energy density in the modulus is equal to $1/2 \, m_X^2 M_{Pl}^2$ and that the radiation energy density is at 
the same level \cite{Acharya:2008bk,Dine:1995kz}.
When $T_{\rm RH}$ is larger than the thermal freeze-out temperature $T_{t.f.o.}$ for this model (the case for $T_{\rm RH} = 100$~GeV in the plot), the temperature 
evolution of $Y_\chi$ is obviously the same as in a standard thermal WIMP framework: $Y_\chi$ follows first the thermal equilibrium distribution along its 
Maxwell-Boltzmann tail, in a phase in which the main source of DM particles is pair production by SM background particles and this is balanced by DM 
pair annihilations, and then at ${T}_{t.f.o.}$, when $n^{eq}_\chi$ becomes smaller than $n_\chi^c$ and pair annihilations become inefficient, $Y_\chi$ settles on 
a constant value.  When $T_{\rm RH}$ is reduced two effects intervene: first of all, the thermal freeze out temperature tends to increase since the modulus  
contribution to the Universe energy density increases $H$ and hence $n_\chi^c$; at the same time, the dominant source of DM particles becomes the modulus 
decays rather than SM pair creation. If $\chi$ number density from the decay exceeds $n_\chi^c$, this source term is balanced by DM pair annihilations and 
$n_\chi$ tracks the quasi-static equilibrium (QSE) density, as defined, e.g., in Ref~\cite{Cheung:2010gj}:
\begin{equation}
  n_\chi^{QSE} \equiv  \left(\frac{B_X \Gamma_X \rho_X}{m_X \langle\sigma_{\rm{eff}} v\rangle}\right)^{1/2}\,.
\end{equation}
For our sample DM model, this is the behavior we find in all cases with large $B_X$ and $T_{\rm RH} < T_{t.f.o.}$: starting at high $T$, $Y_\chi$ follows first 
$Y_\chi^{eq}$, then it becomes equal to $Y_\chi^{QSE}$ up to about $T_{\rm RH}$ when the modulus DM source drops exponentially, $Y_\chi^{QSE}$ crosses 
$Y_\chi^c$ and hence $Y_\chi$ gets frozen. Regarding the temperature scalings in the plot, in the phase when the modulus dominates the energy density and is 
the main entropy source, we see that both $Y_\chi^{QSE}$ and $Y_\chi^c$ are proportional to $T$, except for a short low temperature phase in the examples
for $T_{\rm RH} = 30$~and~6~MeV during which the entropy injected but the modulus decay is still 
negligible compared to the initial entropy and hence $a \propto T^{-1}$, making $Y_\chi^{QSE}$ and $Y_\chi^c$ rise as $T^{-3/2}$. 
For small $B_X$, $Y_\chi^{QSE}$ becomes smaller than $Y_\chi^c$, DM annihilations are inefficient and $n_\chi$ simply scales as 
$B_X \Gamma_X  \rho_X/m_X \cdot t$, up to the reheating temperature when the modulus source drops and $Y_\chi$ becomes constant; for what concerns the
behavior in temperature, once again, in the phase in which the decay injects DM particles, the scaling just given translates into $Y_\chi \propto T$, while for very 
small $B_X$ one can also see a transient in which the amount of DM produced in the decay is small compared to the thermal component and $Y_\chi$ simply 
reflects the entropy increase, decreasing faster than $T$.   

In the example displayed, the specific set of initial conditions implemented to solve the system of equations has a negligible impact on the final comoving 
density of DM particles. In fact the latter is insensitive to the choice of the initial energy density in the moduli and the relative weight with respect to the 
initial radiation energy density provided that the physical mechanism determining the DM relic density starts becoming efficient at temperatures lower than 
the temperature at which the scaling $T \sim {a}^{-3/8}$ begins. More precisely: in all cases considered in this paper, the DM pair annihilation rate
is large enough to guarantee, even in the non-standard cosmological scenarios considered here, chemical equilibrium at $T \gtrsim {m}_{\chi}$;
the final relic densities is then determined by the physics taking place between the thermal freeze-out temperature and the reheating temperature. 
If the $T \sim {a}^{-3/8}$ scaling starts sufficiently earlier than $T_{\rm RH}$, the entropy production guarantees the suppression 
of the DM thermal component and, at the same time, variations in the entropy release with the field energy density are compensated by a different 
efficiency in the non thermal production, leaving then the final result unchanged.  If, on the contrary, the DM thermal relic component is not totally diluted,
the $T \sim {a}^{-3/8}$ phase needs to start before the thermal freeze-out temperature, otherwise the variation of dilution due to entropy release stemming
from the initial conditions has a direct impact on the relic density as well. 

\begin{figure}[t]
 \begin{minipage}[htb]{8cm}
   \includegraphics[width=8 cm, height= 6 cm, angle=360]{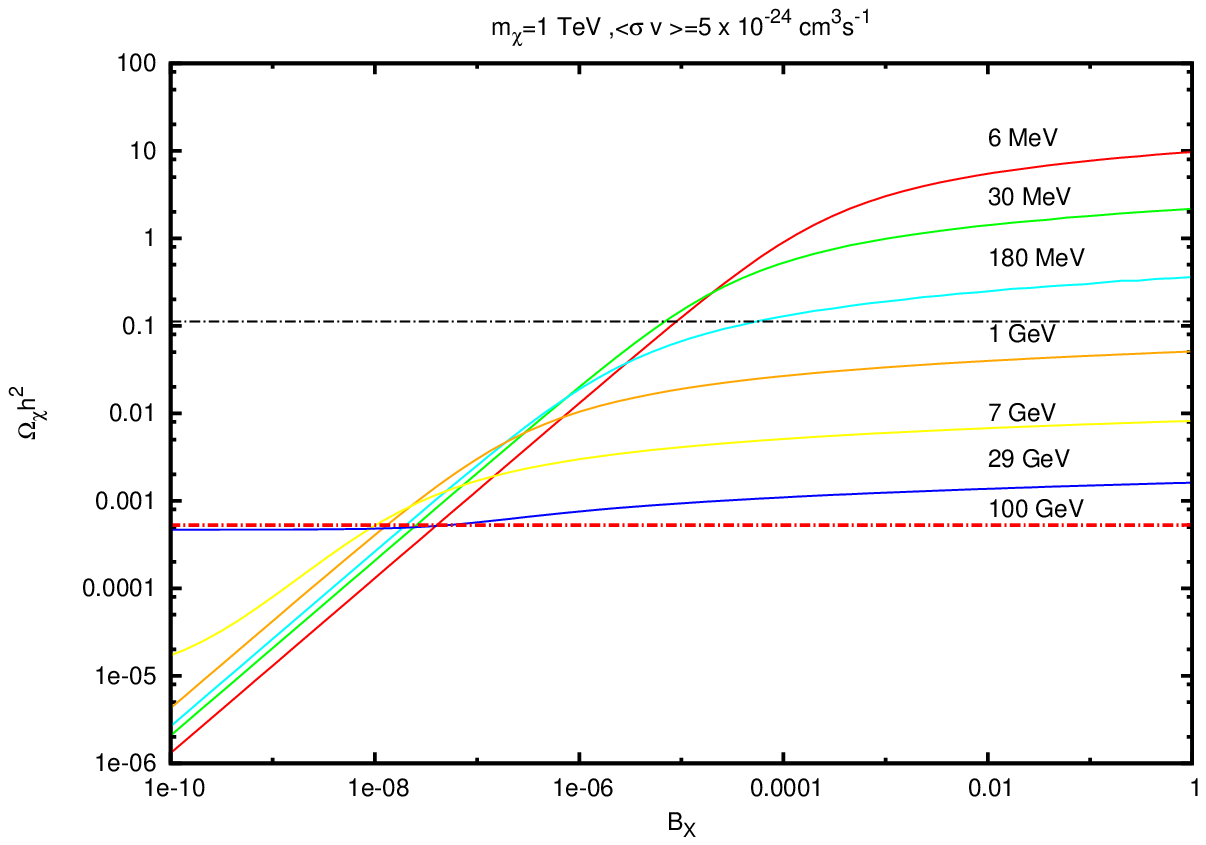}
 \end{minipage}
 \ \hspace{3mm} \
 \begin{minipage}[htb]{8cm}
   \includegraphics[width=8cm, height= 6 cm, angle=360]{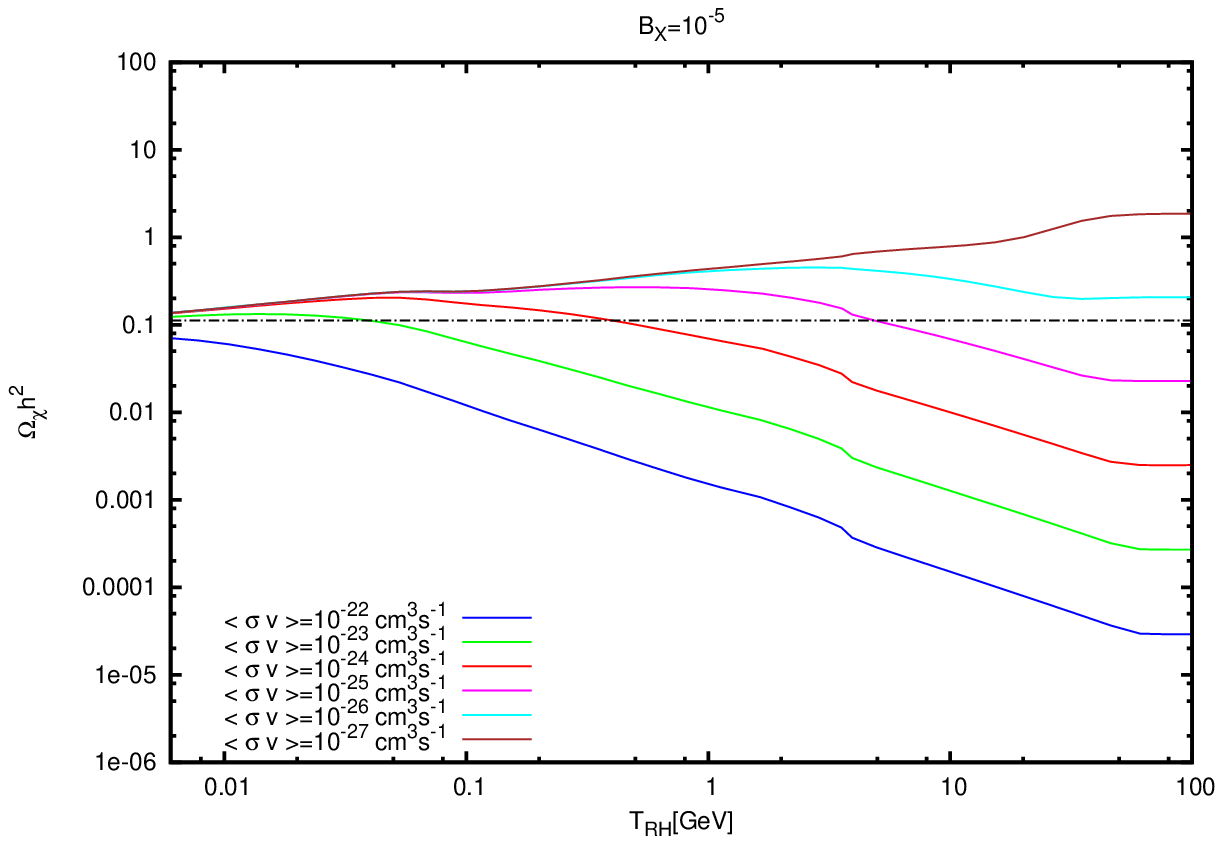}
 \end{minipage}
 \caption{ Left panel: scaling of the relic density with the parameter $B_X$ for the same sample DM model selected for Fig~\protect{\ref{fig:fig1}} and for a few 
  values of ${T}_{RH}$. Right panel: relic density versus ${T}_{RH}$, having fixed $B_X = 10^{-5}$ and rescaling the value of $\langle \sigma v \rangle$.  
  The black horizontal lines represent the cosmological DM density as extrapolated by the WMAP 7-year data~\protect{\cite{Komatsu:2010fb}}.}
 \label{fig:fig2}
\end{figure} 

In Fig.~\ref{fig:fig2} we plot the relic densities for the $\chi$ state. In the left panel we refer to the same model introduced for Fig.~\ref{fig:fig1}, select a few values for 
the reheating temperature and display results as a function of $B_X$; as expected from the discussion above, one can see that $\Omega_\chi$ becomes essentially
independent of $B_X$ in the limit of large $B_X$, while it scales linearly with $B_X$ when the modulus source function is too small to make $n_\chi$ exceeds 
$n_\chi^c$. Also visible at large $B_X$ is the scaling of the relic density with the inverse of $T_{\rm RH}$, as expected from the analytical estimate in Eq.~(\ref{eq:ntc}).
When $B_X$ is small  $\Omega_\chi$ is expected to scale with $T_{\rm RH}/m_X$. In our approach $T_{\rm RH}$ and $m_X$ are correlated; from the instantaneous approximation, Eq.~(\ref{eq:trh}), one expects ${m}_{X}\propto T_{\rm RH}^{2/3}$ giving ${\Omega}_{\chi} \propto T_{\rm RH}^{1/3}$, which is approximately the scaling seen in
the plot for very small $B_X$. 
The dependence on the reheating temperature is shown more explicitly in the right panel of Fig.~\ref{fig:fig2}, where, having fixed $B_X$ to an 
intermediate value, we let the annihilation cross section vary of a few orders orders of magnitude around the value chosen for the plot on the left hand side; the relic density 
scales with the inverse of $\langle \sigma v \rangle$ whenever reannihilation takes place, while evidently the solution does not depend on  $\langle \sigma v \rangle$
in case annihilation processes are inefficient.

\begin{figure}[t]
 \begin{minipage}[htb]{8cm}
   \includegraphics[width=8 cm, height= 6 cm, angle=360]{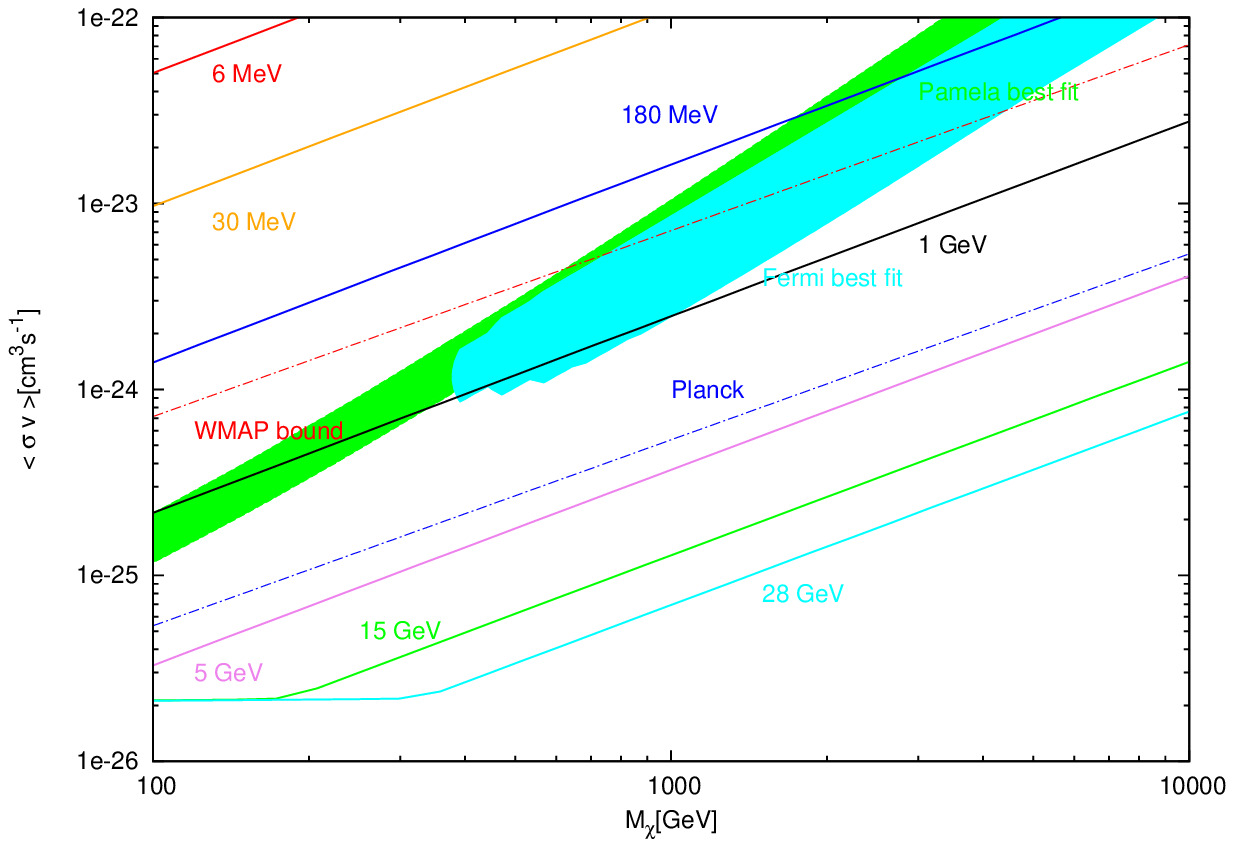}
 \end{minipage}
 \ \hspace{3mm} \
 \begin{minipage}[htb]{8cm}
   \includegraphics[width=8cm, height= 6 cm, angle=360]{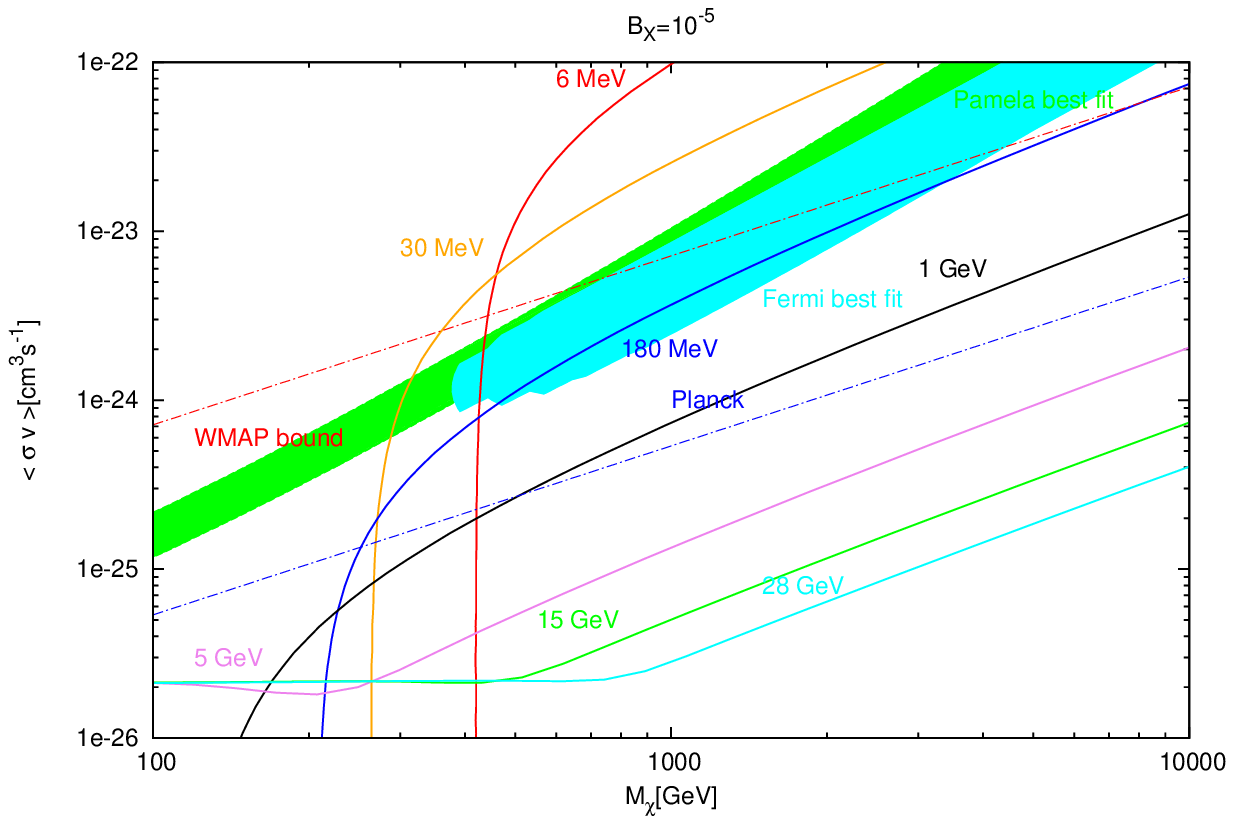}
 \end{minipage}
 \caption{Values of $m_X$ and $\langle \sigma v \rangle$ for which the relic density of the $\chi$ particles matches the cosmological DM density, for a few 
 values of $T_{\rm RH}$ and two representative cases for $B_X$. Also shown are the region in this plane compatible with the Pamela and Fermi electron/positron 
 excesses in case of leptophilic DM~\protect{\cite{Grasso:2009ma}}, and the WMAP limit and Planck projected sensitivity~\protect{\cite{Galli:2009zc, Iocco:2009ch}} 
 stemming from the impact of residual DM annihilations on reionization.}   
 \label{fig:fig3}
\end{figure} 

In most scenarios containing cosmological moduli it is hard to tune the model in such a way that very tiny $B_X$ are obtained, hence the framework we are
discussing becomes interesting mainly when $\chi$ is associated to a large annihilation cross section, preventing the overproduction of DM with respect to the 
experimental bound.  DM models with a $\langle \sigma v \rangle$ which is two or three orders of magnitudes larger than in the standard thermal relic scenario
would be very interesting also from the point of view of indirect DM detection and have been invoked to address the excess in the lepton cosmic ray flux by
Pamela and Fermi. In Fig.~\ref{fig:fig3}, choosing a few sample values of $T_{\rm RH}$ and two representative cases for $B_X$, we scan the parameter space 
$(\langle \sigma v \rangle$~--~$m_\chi)$ searching for configurations in which the $\chi$ relic density matches the central value for the cosmological DM density as 
estimated from the WMAP 7-year data, namely $\Omega_\chi h^2 = 0.1123 \pm 0.0035$ \cite{Komatsu:2010fb}.
A curve corresponding to a given $T_{\rm RH}$ becomes
horizontal when $T_{\rm RH}$ becomes larger than ${T}_{s.f.o.}$, i.e. we recover the standard thermal result of the relic density being independent of mass for
S-wave annihilations; on the other hand it becomes vertical when annihilations become inefficient and hence $\Omega_\chi$ stops depending on 
$\langle \sigma v \rangle$.  In general going from large to small $B_X$, keeping $T_{\rm RH}$ fixed, shifts the results to smaller $\langle \sigma v \rangle$  
and larger $m_\chi$. In the same plot, supposing we are now referring to a leptophilic DM candidate, namely annihilating democratically into the 
three lepton species \cite{Grasso:2009ma}, we have superimposed the region in the parameter 
space which have been found to be compatible with the Pamela and Fermi electron and positron data, as derived, e.g., in Ref.~\cite{Grasso:2009ma}; 
the comparison is meant to be qualitative since we are not considering here a detailed particle physics
scenario, it shows however what are the main trends that should be fullfilled to find an agreement. Also shown is the bounds on leptophilic models following from
WMAP CMB data \cite{Galli:2009zc, Iocco:2009ch}: the limit stems from the impact of residual (namely much later than thermal decoupling) pair annihilations
on reionization, and will be soon improved by the Planck experiment in case of no signal.

The last issue we wish to discuss in this Section is an implicit dependence on the modulus mass $m_X$ we have ignored so far: As mentioned above 
we have been varying $m_X$ to retrieve different values of $T_{\rm RH}$ as extrapolated from the numerical solution of the system of coupled equations; the 
underlying assumption here is that we computed the modulus decay assuming gravitational coupling and a two body final state. Having in mind more
general scenarios like those, e.g., in Ref.~\cite{Nakamura:2006uc,Dine:2006ii,Kohri:2004qu,Moroi:1999zb,Endo:2006zj,Moroi:1994rs}, 
we may consider replacing:
\begin{equation}
 \Gamma_X =  \frac{1}{4 \pi} \frac{m_X^3}{M_{Pl}^2} \quad \rightarrow \quad \Gamma_X=\frac{m_X^3}{\Lambda_{eff}^2}\,,
\end{equation}
where now ${\Lambda}_{eff}$ encodes both the coupling of the effective operator responsible for the decay and the kinematical factors. From the 
approximation of instantaneous reheating one sees that, to keep $T_{\rm RH}$ fixed after this replacement, one needs simply to approximately shift:
\begin{equation}
 m_X  \quad \rightarrow \quad  m_X \cdot \left(\frac{\Lambda_{eff}^2}{4 \pi M_{Pl}^2}\right)^{1/3}\,.
\end{equation}
The modulus mass however appears explicitly also in Eq.~(\ref{eq:dm}) when, in the DM source function from modulus decays, one converts 
from the modulus energy density to its number density. To compensate for this and use results displayed in this  and the next Sections, one then 
should also shift the values reference values for $B_X$ as:
\begin{equation}
 B_X \quad \rightarrow \quad  B_X \cdot \left(\frac{\Lambda_{eff}^2}{4 \pi M_{Pl}^2}\right)^{1/3}\,.
\end{equation}

\section{Impact of a toy model on the MSSM}

In this Section, while still referring to the schematic picture with a single cosmological modulus $X$ parametrized through its decay width $\Gamma_X$ and the 
DM yield $B_X$, we introduce an explicit particle physics scenario for the $\{{\chi_a}\}$ fields, considering neutralino DM in the MSSM. As already mentioned, 
a scenario with DM production from cosmological moduli can arise quite naturally in supergravity/superstrings theories. Some scenarios, such as gauge mediated 
supersymmetry breaking are actually troublesome since the moduli tend to be light and decay after the onset of BBN, see, 
e.g.~\cite{Lyth:1995ka,Asaka:1997rv,Acharya:2009zt,Acharya:2010af}; to solve this problem, one needs to invoke a mechanism of dilution of the moduli 
number density, making these models not viable for non-thermal DM production.

In the MSSM there are four neutralinos, spin 1/2 Majorana fermions obtained as the superposition of two neutral gauginos, the Bino $\tilde B$ and the Wino 
$\tilde W^3$,  and the two neutral Higgsinos $\tilde H^0_1$ and $\tilde H^0_2$:
\begin{equation}
 {\tilde{\chi}}_{i}={N}_{i1}\tilde{B}+{N}_{i2}\tilde{W}+{N}_{i3}{\tilde{H}}_{1}^{0}+{N}_{i4}{\tilde{H}}_{2}^{0}\,,
\end{equation}
where the coefficients $N_{ij}$ are the elements of the matrix which diagonalizes the neutralino mass matrix, and are mainly a function of the Bino and the Wino mass 
parameters $M_1$ and $M_2$, and of the Higgs superfield parameter $\mu$, while depend rather weakly on $\tan\beta$, the ratio of the vacuum expectation values 
of the two neutral components of the SU(2) Higgs doublets, which appears in the off-diagonal terms of neutralino mass matrix. The hierarchy between $M_1$, $M_2$
and $\mu$ sets hence whether the lightest neutralino, which we are assuming also as lightest supersymmetric particle (LSP) and stable, 
is mostly bino-, wino- or Higgsino-like.  
From the point of view of DM production in the early Universe, pure Binos have pair annihilation cross sections dominated by SM fermion final states, which are helicity 
suppressed (S-wave annihilation cross sections proportional to the square of the mass of the final state fermion) and scale approximately with the inverse of the forth
power of the mass of the corresponding sfermions; given current accelerator bounds on sfermion masses, Binos tend in general to have a too large annihilation 
cross section to be thermal DM relics. On the other hand Winos and Higgsinos have unsuppressed annihilation cross sections into $W$ bosons and their
annihilation cross section tends to be too large for thermal production unless one considers heavy states, about 2.4 and 1.1 TeV, respectively, 
for a pure Wino and a pure Higgsino, since, in this case, the annihilation cross sections scales approximately with the inverse 
of the square of the neutralino mass (we will show results obtained computing tree-level annihilation amplitudes; in case of TeV Winos and Higgsinos 
actually the result would change slightly, shifting the masses to slightly larger values, when taking into account that, for such heavy states, the weak interaction 
acts as a long-range attractive force which deforms the wave function of the annihilating DM pair, an effect usually referred as Sommerfeld enhancement,
see, e.g., \cite{Hryczuk:2010zi} and references therein).  Since we will be mainly interested in discussing the shift on the mass scale for neutralino DM due 
to non-thermal effects, we refer here to a supersymmetric framework maximizing this dichotomy between underproduced and overproduced thermal states, 
the so-called "Split Supersymmetry"
scenario~\cite{ArkaniHamed:2004fb,Giudice:2004tc}. This indicates a generic supersymmetric extension to the SM in which fermionic superpartners have
a low mass spectrum (say at the TeV scale or lower), while scalar superpartners are heavy, with a mass scale which can in principle range from hundreds of TeV 
up to the GUT or the Planck scale \cite{ArkaniHamed:2004fb}, a feature which can occur in a wide class of theories, see, 
e.g.~\cite{ArkaniHamed:2004yi,Antoniadis:2004dt,Kors:2004hz}. Leaving out of the discussion also gravitinos which are assumed to be heavy and not produced in 
the modulus decay, the system $\{{\chi_a}\}$ reduces to neutralinos and charginos, whose annihilation and coannihilation effects we treat interfacing the model
the \ds\ package \cite{Gondolo:2004sc}. Finally for what regards the Higgs sector, the scenario has only one light state SM-like Higgs; the value of its mass, 
as well as $\tan\beta$ have no sizable impact on the overall picture, hence we keep them fixed to sample values, respectively, $114.4\,\, \mbox{GeV}$ and $10$. 

\begin{figure}[t]
 \begin{minipage}[htb]{8cm}
   \includegraphics[width=8 cm, height= 6 cm, angle=360]{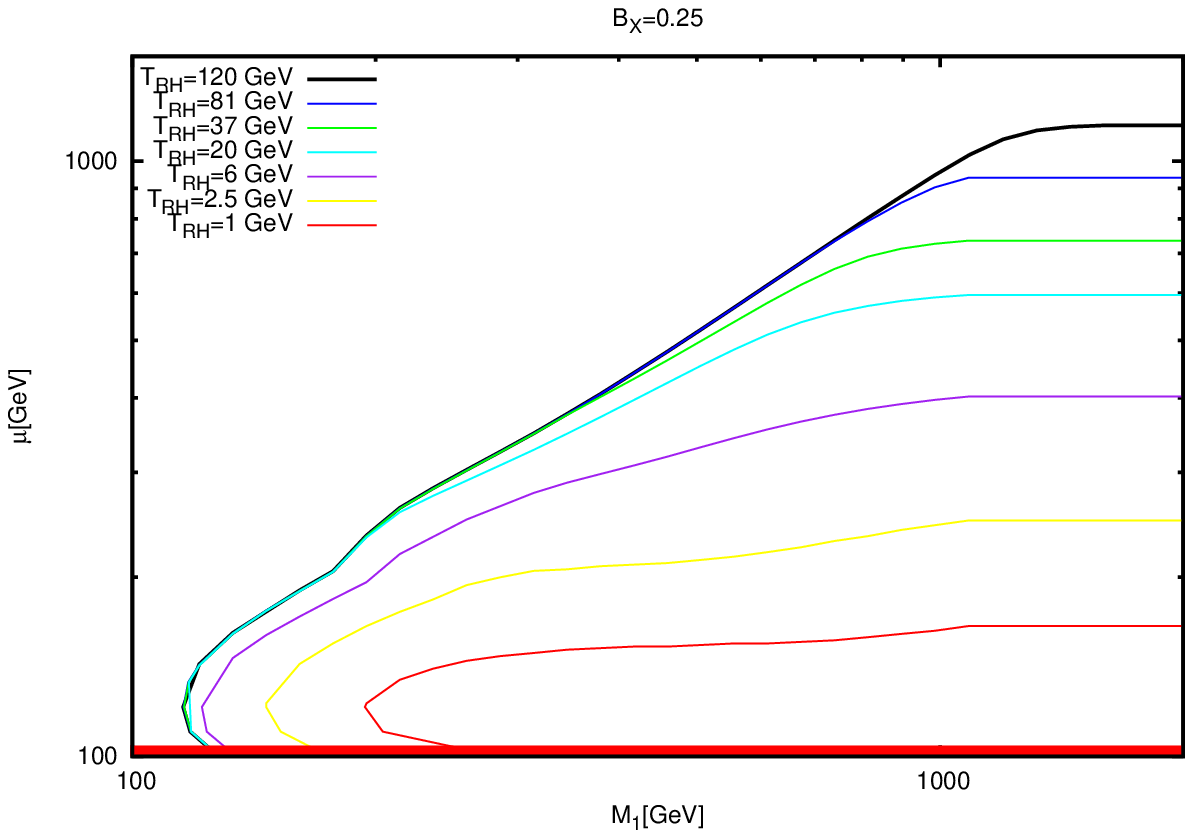}
 \end{minipage}
 \ \hspace{3mm} \
 \begin{minipage}[htb]{8cm}
   \includegraphics[width=8cm, height= 6 cm, angle=360]{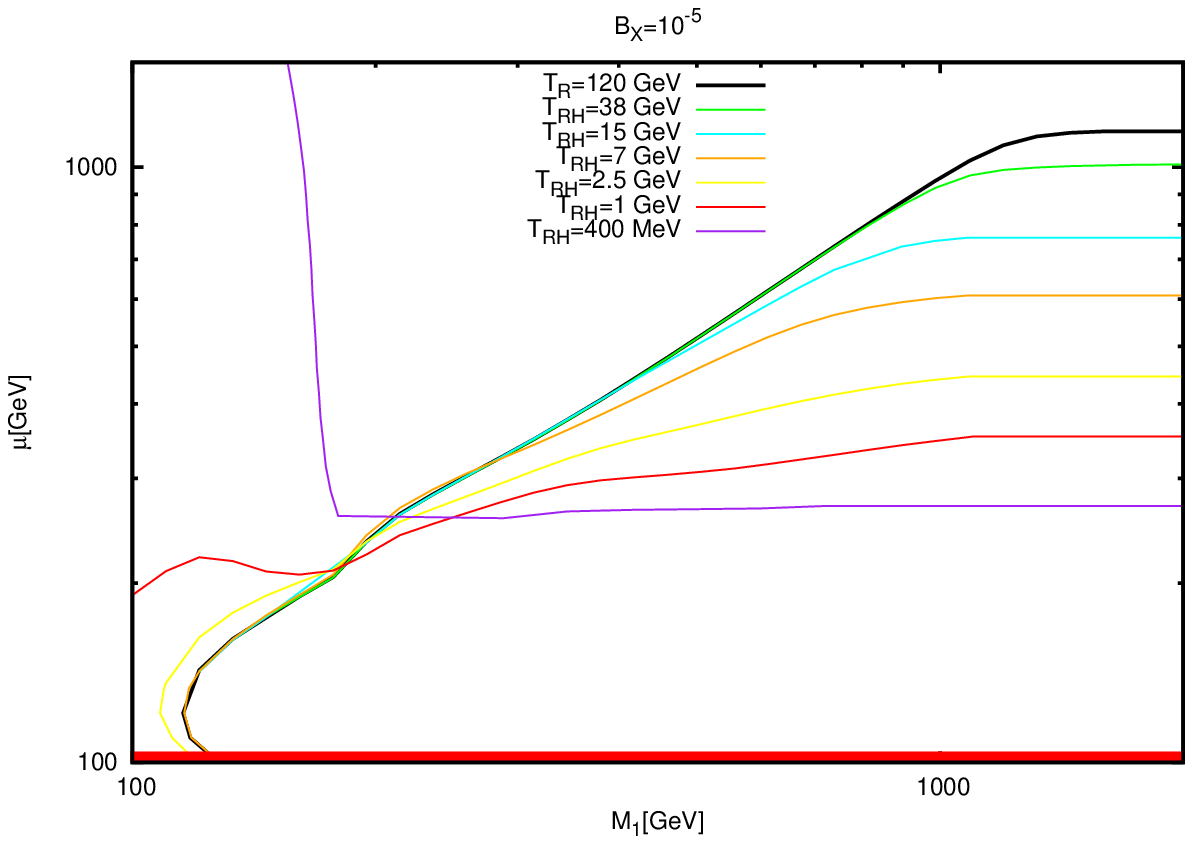}
 \end{minipage} \\
 \begin{minipage}[htb]{8cm}
   \includegraphics[width=8 cm, height= 6 cm, angle=360]{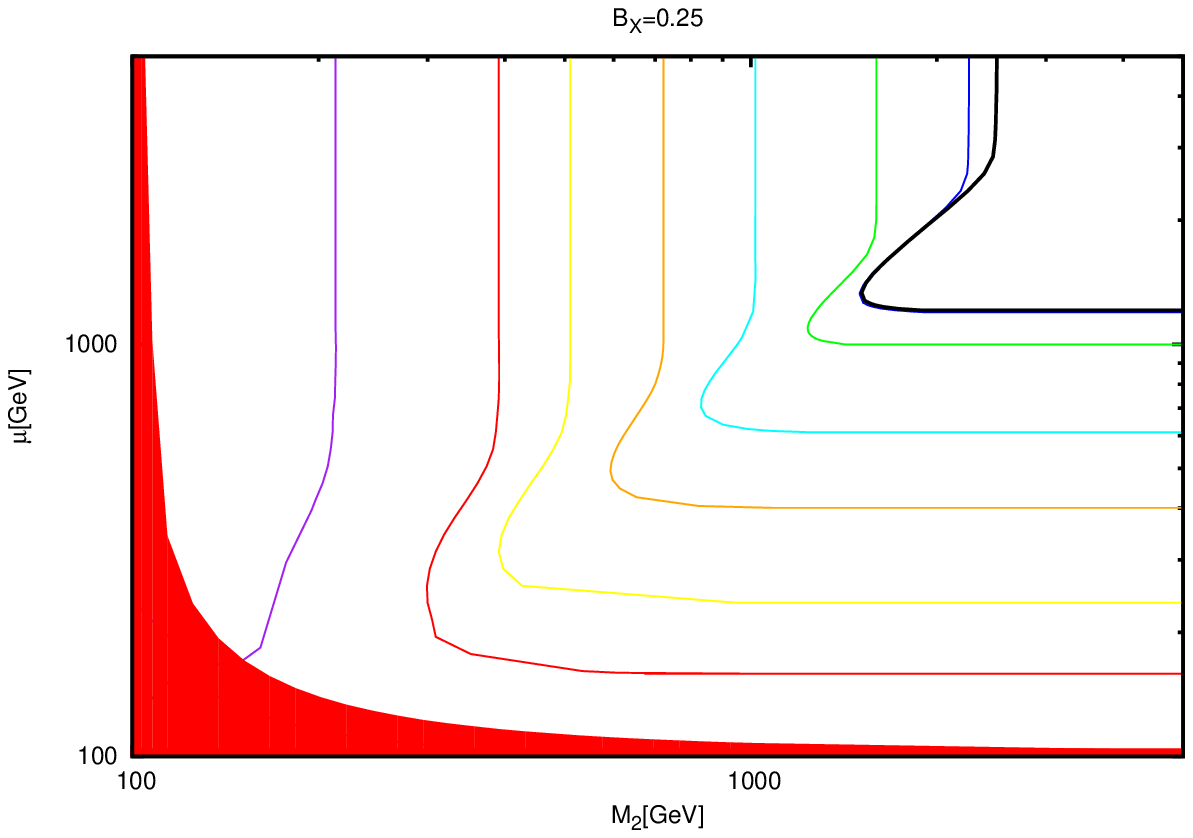}
 \end{minipage}
 \ \hspace{3mm} \
 \begin{minipage}[htb]{8cm}
   \includegraphics[width=8cm, height= 6 cm, angle=360]{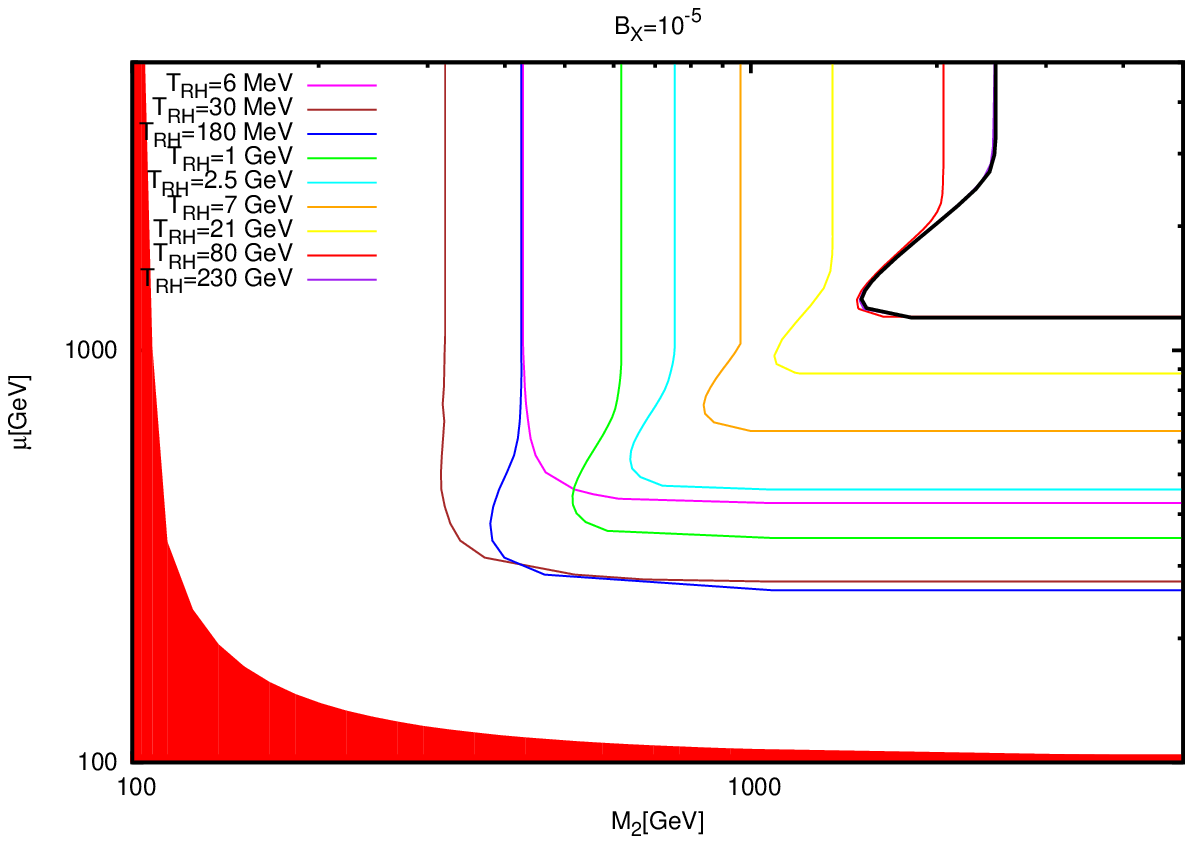}
 \end{minipage}\\
 \begin{minipage}[htb]{8cm}
   \includegraphics[width=8 cm, height= 6 cm, angle=360]{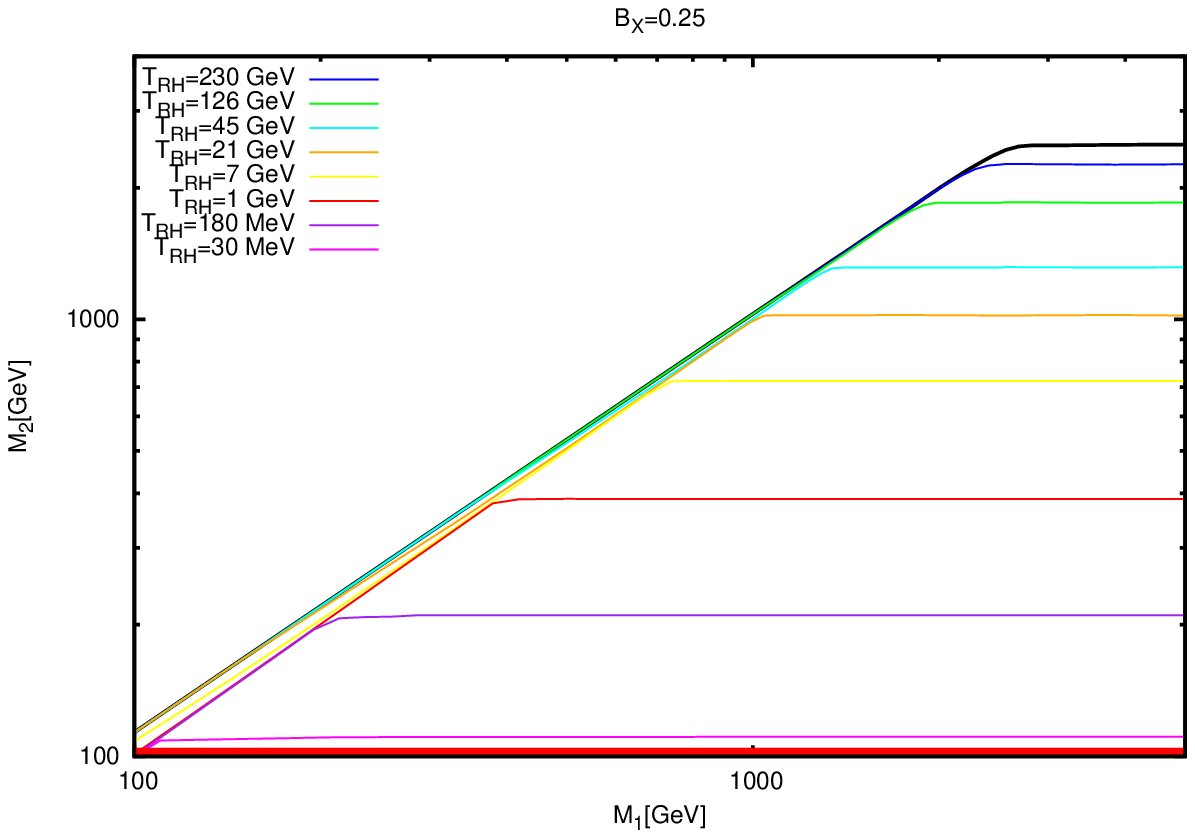}
 \end{minipage}
 \ \hspace{3mm} \
 \begin{minipage}[htb]{8cm}
   \includegraphics[width=8cm, height= 6 cm, angle=360]{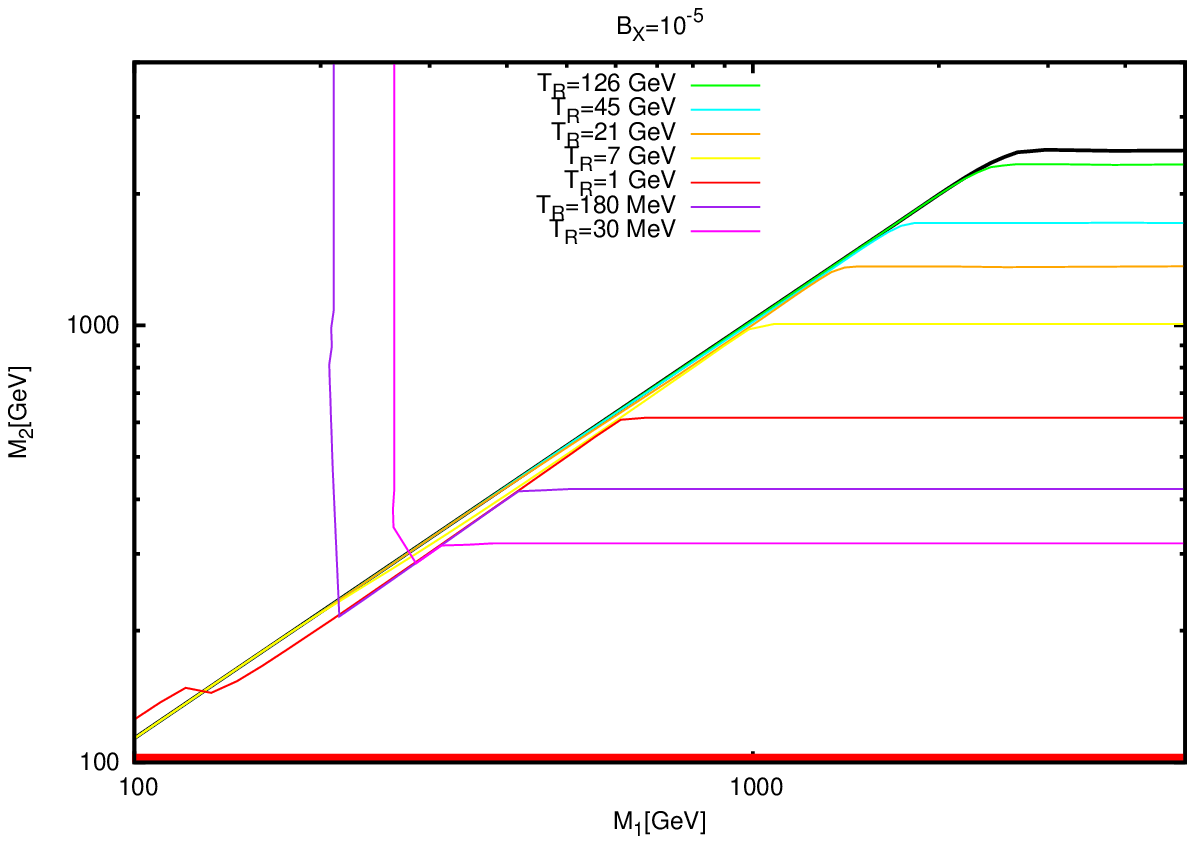}
 \end{minipage}
 \caption{Models with relic density equal to the central value from the 7-year WMAP data analysis, i.e. $\Omega_\chi h^2 = 0.1123$, in 
  two-dimensional slices of the $M_1$, $M_2$, $\mu$ parameter space, corresponding to the limit in which the third parameter is heavy (set to 10~TeV in the 
  numerical  computation), for a few values of $T_{\rm RH}$ as indicated in the plots, and two values of $B_X$, namely 0.25 for left panels and $10^{-5}$ for 
  right panels. The filled areas correspond to regions violating the LEP lower bound on the chargino mass.}
 \label{fig:fig4}
\end{figure}

In Fig.~\ref{fig:fig4} we scan the parameter space $M_1$, $M_2$ and $\mu$ searching for models whose relic abundance matches the central value from
the 7-year WMAP estimate of the DM density in the Universe. There are three pairs of plot in which we vary two of the parameters, fixing the third to a heavy scale;
in each pair, one plot is for a large $B_X$, while the other is for a small but not negligible $B_X$. As in the previous Section, we vary $m_X$ to change the reheating
temperature scale, assuming a two-body gravitational decay for the modulus. The thick black solid line corresponds, in each plot, to a reheating temperature exceeding 
the thermal freeze-out temperature for all models along the curve, namely it gives the models matching the cosmological DM density we would also obtain
in the standard picture without non-thermal DM sources: In the $M_1$-$\mu$ plane this happens, starting at small neutralino masses, close to the diagonal 
$M_1=\mu$ since it requires a tuning of the right amount of Higgsino and Bino component in the LSP, suppressing the large Higgsino annihilation cross section 
with the Bino one, which in Split SUSY is extremely small. Would we have allowed for lighter sfermions and other light Higgs states, this curve would have moved
only slightly further away for the diagonal, except when sfermion coannihilations or S-channel resonant annihilations on a Higgs take over in setting the 
effective thermally averaged annihilation cross section, as happens, e.g., in portions of the mSUGRA parameter space, see, 
e.g.~\cite{Edsjo:2003us} -- we will not discuss these exceptions here. As already mentioned, at about 1.1~TeV 
a pure Higgsino saturates the thermal relic density bound. Turning to the $M_2$-$\mu$ plot, Winos have an even larger annihilation cross section than Higgsinos
and the thermal relic density curve just goes from a pure Higgsino to a heavier pure Wino through a transient with large Higgsino-Wino mixing. Finally the behavior 
in the $M_1$-$M_2$ plane is more peculiar since from the structure of the neutralino mass matrix, Bino and Wino do not mix and, below the mass scale for a pure
Wino thermal relic candidate, the tuning here is between the mass spitting between the Bino LSP and the second lightest neutralino and the lightest chargino, 
which are Wino-like and and whose coannihilations in the early Universe set the thermal relic abundance of a Bino LSP (there are chargino and neutralino 
coannihilations even for pure Winos and Higgsinos, but with less dramatic effects). Turning on the non-thermal component from the modulus decays, when $B_X$ 
is large, essentially one just sees in the plots the scaling sketched in Eq.~(\ref{eq:ntc}), with Higgsinos and Winos saturating the WMAP preferred value 
for $\Omega_\chi$ with a progressively larger $\langle \sigma v \rangle$ as $T_{\rm RH}$ decreases, and hence for a progressively smaller LSP mass, 
covering the whole parameter space for Higgsinos lighter than 1.1~TeV and Winos lighter than 2.4~TeV (as the Higgsino mass approaches the $W$ boson 
threshold the cross section stops increasing; this explains
the shape of isolevel curves in that region). 
A detection at an accelerator of such LSP configurations, hopefully combined with a DM detection signal,
would indeed be an indication of a non-standard cosmological phase at DM generation, with non-thermal production as primary scenario (there would also be 
further possibilities, such as, e.g., the increase of the Universe expansion rate at freeze-out induced by a quintessence 
component~\cite{Salati:2002md,Profumo:2003hq} or a modification of the gravity theory~\cite{Catena:2004ba}).  
When $B_X$ is small, there is a smooth transition from 
the regions where the scaling in Eq.~(\ref{eq:ntc}) applies to those where annihilations stop playing a role and Eq.~(\ref{eq:ntnc}) applies instead; the latter makes 
even pure Binos, which, we underline again, in our sample MSSM setup have extremely small annihilation cross sections, become cosmologically viable, another 
configuration which, if singled out at accelerator and/or DM searches, would point to a non-standard early Universe cosmological history 
(in plots the filled region stands for the region in which the LEP bound on the chargino mass $m_{\chi^+}>103.5$~GeV is violated; 
Tevatron and the recent LHC constraints, such as~\cite{Khachatryan:2011tk,Aad:2011xm}, are not shown since we have made just 
schematic assumptions on sfermions and not discussed at all gluinos, the particles most critical for a early discovery at a hadron collider).
     
\section{Non-thermal DM production in the G2-MSSM}

As an example of framework in which we can make definite predictions for the spectra of both the set of particles $\{{\chi_a}\}$ and the cosmological moduli, 
we discuss the case of the G2-MSSM~\cite{Acharya:2007rc,Acharya:2008zi}. 
Within a specific class of string M-theories, in this scenario the compactification of 
extradimensions gives rise to a $N=1$ supergravity in which SUSY breaking, due to the dynamics of a hidden sector, is trasmitted to the visible sector 
by a combination of gravity (dominant contribution) and anomaly mediation. We briefly summarize here the main features of the spectrum, following 
Ref.~\cite{Acharya:2008zi}: In the G2-MSSM the visible sector can be described by a GUT theory broken into the MSSM at the unification scale ${M}_{\rm unif}$, 
at about $10^{16}$~GeV, coinciding with the compactification scale. The RGEs boundary conditions are mainly functions of the gravitino mass ${m}_{3/2}$, 
which can be estimated from the UV theory parameters to lie in the range between ten and several hundred TeVs. The gauginos are expected to be the 
lightest SUSY particles; at ${M}_{\rm unif}$ the gaugino masses are generated from a universal loop-suppressed gravity mediation contribution combined
with a non-universal anomaly mediation term. The ratio of the gaugino masses to the gravitino mass depends almost linearly on the quantity
$\delta$ that parametrizes a threshold correction to the unified gauge coupling; in the following, $\delta$ will be kept as a free parameter. The value of the masses at the 
electroweak scale is computed following the RGE evolution, including threshold corrections, such as the very large correction coming from higgs-higgisino loops, 
which is proportional to $\mu$ \cite{Pierce:1996zz}; whether the lightest neutralino is the Bino or the Wino depends on the sign and magnitude of this latter 
correction. For $\mu>0$ and for $\delta$ in the range $-10 \le \delta \le 0$ the lightest neutralino is a pure Wino, with mass in the range between about 100 
and few hundred GeV (even the gluino is fairly light, $m_{\tilde{g}}<1$~TeV, a feature implying a rather rich phenomenology at LHC 
\cite{Feng:1999fu,Acharya:2009gb,Feldman:2010uv} and making the model testable in the near future). Concerning the other states in the MSSM spectrum, 
the Higgsino mass parameter $\mu$ and soft SUSY breaking term $B\mu$ are generated by a Giudice-Masiero mechanism and are heavy, of the order of $m_{3/2}$. Sfermions are also heavy with a flavor universal contribution to their soft masses at ${M}_{\rm unif}$ being about ${m}_{3/2}$; RGEs affect mostly the third generation 
of squarks with the stops and the left handed sbottom becoming the lightest sfermions at the Electroweak scale (the left-handed stop mass becomes about 
$0.25 \cdot {m}_{3/2}$, the right handed stop and the left handed sbottom masses about $0.5 \cdot {m}_{3/2}$). In the Higgs sector $\tan \beta$ is fixed 
by $\mu$ and $B\mu$ through the electroweak symmetry breaking condition and takes a value of order one; the light CP even Higgs is Standard Model like, while 
all the other Higgs bosons are heavy, again at about the ${m}_{3/2}$ scale. In summary, the features relevant to discuss non-thermal DM production in 
this model are: The pure Wino LSP as DM candidate, as enforced by the proper choice of $\delta$ (we will restrict to values $\delta < -3$ since they are theoretically 
favored~\cite{Acharya:2008zi}) and required to provide an annihilation cross section sufficiently large for the model to fit into the scenario in which the branching 
ratio for the decay of the moduli into the LSP is unsuppressed; A charged Wino as next to lightest SUSY particle, with a tiny mass splitting with respect to the 
LSP, about 200 MeV, due to the one-loop electroweak corrections to the neutralino and chargino masses~\cite{Feng:1999fu}; The possibility for the 
moduli to decay into gluinos and third-generation squarks. While other SUSY particles do not play a role in our analysis, the relevant part of the spectrum 
will be computed here implementing the appropriate one-loop RGE running.

For what regards the moduli fields,  as already mentioned, in a string framework like the G2-MSSM, they arise in the effective supergravity theory after the 
compactification of the extra-dimensions. The theory predicts the presence of a large number of moduli fields with mass of order or heavier than the gravitino 
mass. In our numerical computation of the DM relic density, we follow the scenario outlined in~Ref.~\cite{Acharya:2008bk}: The modulus sector is composed by 
$N+1$ fields with $N=\mathcal{O}(50-100)$ including:
\begin{itemize}
\item 1 heavy modulus $X_N$ with  $m_{X_N}=600 \, m_{3/2}$;
\item 1 meson field $\Phi$ with $m_\Phi =1.96 \, {m}_{3/2}$;
\item N-1 light moduli ${X}_{i}$ with $m_{X_i}=1.96 \, {m}_{3/2}$. 
\end{itemize}
Their decay rates can be written in the form given in Eq.~(\ref{eq:dw}), i.e. it is proportional to the mass of the modulus to the third power and inversely proportional 
to the square of the reduced Planck mass. The constant in front can be computed esplicitly \cite{Acharya:2008bk} in the model; we keep $D_{X_N}$ 
for the heavy modulus and $D_\Phi$ for the meson fixed to the benchmark values of, respectively, 2 and 710, while ${D}_{{X}_{i}}$, which is one the quantities the 
LSP relic density is mostly sensitive to, will be treated as a free parameter allowed to vary in the range between 4 and 16 (preferred range in the 
scenario considered here~\cite{Acharya:2008bk}). 
With this choice, the lifetimes for the three type of states is split to about: $10^{-10}$~s for $X_N$, $10^{-5}$~s for 
$\Phi$ and $10^{-3}-10^{-2}$~s for the light moduli. The branching ratios into SUSY particles of the decays are also calculable in this theory, with the main
channel being into squark pairs, mostly the lightest stop, which in turn cascade down to the LSP and the chargino; in general, the branching ratio of decay 
of the light moduli into Susy particles is 25\% with on average two DM particles produced at the end of the decay chain, giving $B_{X_i} \sim 0.5$.
Gravitinos are produced in the heavy modulus decay, while gravitino pair production in the decay of the meson and the light moduli is kinematically forbidden
for the given values of ${m}_{\phi}$ and ${m}_{{X}_{i}}$, a choice quite natural for the G2-MSSM but still not totally general~\cite{Acharya:2007rc}. 
We will comment further on this point below. Gravitinos are also long lived:
\begin{equation}
\Gamma_{3/2}=\frac{1}{288 \pi}\frac{m_{3/2}^3}{M_{Pl}^2}
\end{equation}
producing one SUSY particle per decay, cascading again into one DM particle.

The system in Eq.~(\ref{eq:system}) is solved numerically for this G2-MSSM setup, with the set of moduli just outlined and having chosen 
$N=99$ and including the gravitino as $\psi$ field. The quantities which are kept as free parameters in our analysis are the gravitino mass, the parameter 
$\delta$ (allowing to shift the ratio between LSP and gravitino mass ) and $D_{X_i}$. Except for gravitinos, all other decay products are, for the moment, 
treated as particles in kinetic equilibrium. The system is evolved starting with the oscillations of the heavy modulus, 
when its initial energy density $1/2 \, m^2_{X_N} M_{Pl}^{2}$ is equal to the radiation energy density, while the other moduli are included in the system 
at the beginning of their oscillations. For what regards the generation of DM, the relevant production phase is only the one from the decays of the light moduli, 
since the thermal DM component, as well as those from the decay of the heavy modulus and the meson, get diluted in the entropy injection phases. 
The dependence of DM comoving number density $Y_\chi$ on temperature in this scenario is perfectly specular to those shown in Fig.~\ref{fig:fig1} for models
whose number density follows first a phase of the quasi-static equilibrium and then reannihilation.  Gravitinos play a marginal role: produced in the heavy 
modulus decay, they get diluted and decay at a late stage (possibly after the end of the reannihilation phase for $\chi$) when $Y_{3/2}$ is tipically 
3 to 4 orders of magnitude smaller than the final $Y_\chi$, hence not contributing significantly to the DM relic density.

\begin{figure}[t]
 \begin{minipage}[htb]{8cm}
   \includegraphics[width=8 cm, height= 6 cm, angle=360]{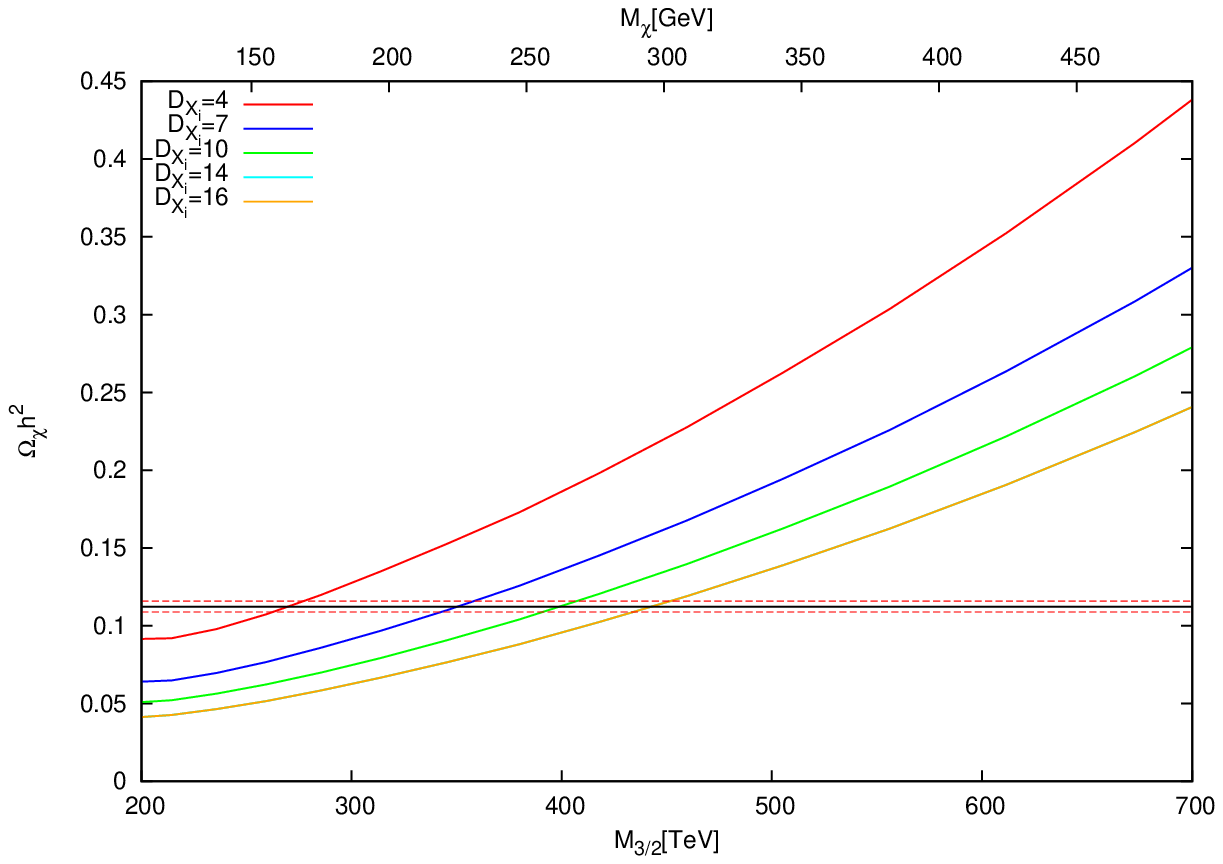}
 \end{minipage}
 \ \hspace{3mm} \
 \begin{minipage}[htb]{8cm}
   \includegraphics[width=8cm, height= 6 cm, angle=360]{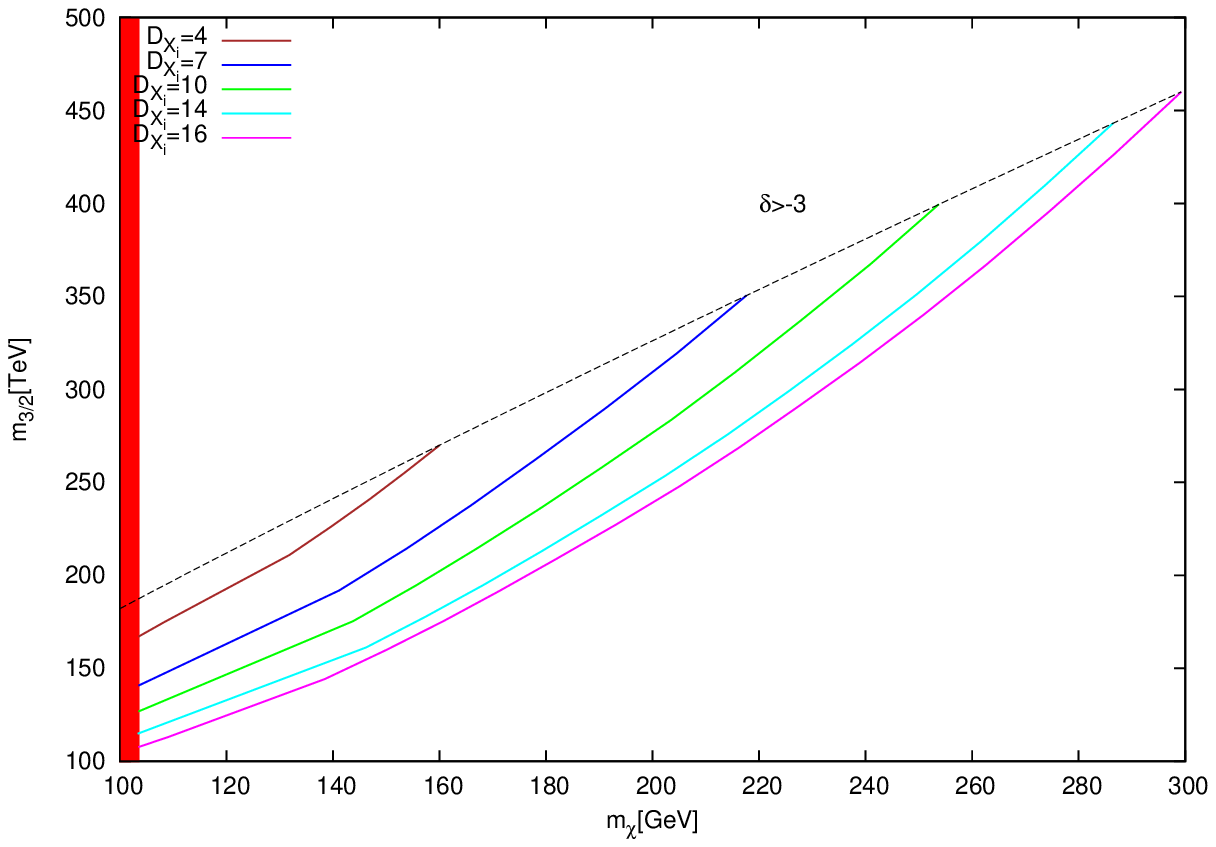}
 \end{minipage} 
 \caption{{\sl Left panel}: Wino relic density versus the gravitino mass $m_{3/2}$ for a few sample values of $D_{X_i}$ and $\delta =-3$; the upper
  horizontal scale shows the corresponding value of for $m_\chi$ for this specific value of  $\delta$. The band gives the $1-\sigma$ determination of the 
  DM relic density from the 7-year WMAP dataset. {\sl Right panel}: models with relic density equal to the central value from the WMAP data, 
  in the plane ${m}_{\chi}$ versus ${m}_{3/2}$, obtained by varying $\delta$ in the range $-10 < \delta < -3$ and few sample values of $D_{X_i}$. The 
  filled area marks the region violating the LEP lower limit on the chargino mass; the region of the plane above the dashed line would correspond
  to models with $\delta > -3$.}
 \label{fig:fig5}
\end{figure}   

In left panel of Fig.~\ref{fig:fig5} we plot the neutralino relic density versus $m_{3/2}$, for a few values of $D_{X_i}$ and a sample value for $\delta$, showing
also on the upper horizontal scale the corresponding value of the Wino LSP mass for such given $\delta$.
 Given that the reannihilation regime applies, from Eq.~(\ref{eq:ntc}) we expect $\Omega_\chi$ to be 
proportional to $m_\chi$ and inversely proportional to $\langle \sigma v \rangle$ and $T_{\rm RH}$, with the latter in turn approximately proportional to 
$D_{X_i}^{1/2}\,(m_{3/2})^{3/2}$, see Eq.~(\ref{eq:trh2}) where the scaling in the modulus masses has been replaced by the scaling in terms of the gravitino 
mass. In the limit in which the Wino pair annihilation cross section just scales with the inverse of the square of the Wino mass, one would find:
\begin{equation}
 \Omega_\chi h^2 \propto \frac{[F(\delta)]^3 \left(m_{3/2}\right)^{3/2}}{D_{X_i}^{1/2}}
\end{equation}
where the function $F(\delta)$ parametrizes the quasi-linear relation between $m_\chi$ and $m_{3/2}$. In the plot, the result of the full numerical 
solution roughly confirms these approximate scalings, except for small $m_\chi$ for which $\langle \sigma v \rangle$ 
is not inversely proportional to $m_\chi^2$. To match the experimental value the DM abundance, lighter $m_{3/2}$ and larger $D_{X_i}$ are favored.
In the right panel of Fig.~\ref{fig:fig5} we consider the plane ${m}_{\chi}$ versus ${m}_{3/2}$ and, varying $\delta$ and for a few values of ${D}_{{X}_{i}}$, 
we plot models that have ${\Omega}_{\chi}{h}^{2}$ equal to the mean value from the WMAP data; the plot illustrates the fact that, even in a model as 
constrained as the G2-MSSM, there is still a rather large sensitivity to the parameters setting the theory at high energy. A relic density compatible 
with cosmological measurements is obtained for LSP lighter than about 300~GeV and for reheating temperatures in the range between about 100~MeV
and 1~GeV.  The results of our analysis are consistent, as an overall picture, with the results presented in Refs.~\cite{Feldman:2010uv,Acharya:2008bk},
although there are slight numerical differences when comparing model by model; most likely these differences stem mainly from the determination of the mass 
spectrum of the G2-MSSM which is probably less accurate in our work, although the more careful numerical treatment implemented here for the relic density
calculation may have some impact as well. As a final remark, we mention that we have also crosschecked the result that, to obtain a relic density compatible 
with the DM density as measured by WMAP,  it is necessary to forbid the decay of the light moduli into gravitinos; in case it is not, in all G2-MSSM setups,
gravitino decays become the main dark matter source, at a stage when reannihilations are inefficient, largely overproducing dark matter.

\section{Kinetic equilibrium and decoupling in the G2-MSSM}

In Section~\ref{sec:general} we have emphasized that Eq.~(\ref{eq:dm}), tracing the evolution of the number density of the $\chi_a$ particles, has been
written assuming that kinetic equilibrium between the $\chi_a$ states and the thermal bath particles is maintained at all stages over which the comoving 
number density changes. Whether this assumption is valid or not depends on the efficiency of the scattering processes on thermal bath particles, an 
issue with is usually addressed invoking crossing symmetry arguments relating the scattering to the annihilation cross section; in most explicit models 
however the two processes are not related via crossing symmetry and one should actually study this problem case by case. We focus here on the G2-MSSM
(a more general framework with non-thermal Wino DM will be also considered at the end)  
and discuss the steps which should be followed when relaxing the hypothesis of kinetic equilibrium, introducing a more general set of Boltzmann equations.

The energy spectrum of SUSY particles produced in the decay of moduli is usually very different from the thermal distribution; in particular in the G2-MSSM
scenario, light particles are generated in the decay of very heavy fields. The cascade process generally starts with the production a pair of squarks, followed
by their decay into gluino and quark, and with the gluino in turn decaying with a three body process into the LSP, the Wino-like chargino or the Bino, with branching
ratios depending on parameters in the model. As a last step, the Bino decays as well into the chargino or the LSP, while the chargino, given the small mass 
splitting with respect to the LSP, has a longer lifetime. The chargino decay occurs either through a two body process in which a pion is produced together with the LSP, 
or through a three-body in which a neutrino and an electron are produced; the rates of these processes are given by, respectively, 
\cite{Chen:1995yu,Chen:1996ap}:
\begin{equation}
  \label{eq:chdecay}
  {\Gamma}_{\chi^\pm\,,{\rm 2b}}=\frac{2{f}_{\pi}^{2}{G}_{\rm F}^{2}}{\pi }{\Delta m_\chi^2}\sqrt{\Delta m_\chi^2-{m}_{\pi}^{2}} 
  \quad \quad {\rm and} \quad \quad 
  {\Gamma}_{\chi^\pm\,,{\rm 3b}}=\frac{2{G}_{\rm F}^{2} \Delta m_\chi^5}{15 {\pi}^{3}}\;,
\end{equation}
where $\Delta m_\chi$ is the chargino-neutralino mass splitting, ${f}_{\pi}=93$~MeV is the pion decay constant and $G_{\rm F}$ 
is the Fermi constant. 
The two-body decay is dominant when kinematically allowed; this is the case in the G2-MSSM, since the minimum mass splitting between charged and neutral Wino, induced but electroweak radiative corrections to the two masses is 
$\Delta m_\chi \simeq 160$~MeV~\cite{Feng:1999fu} .
We have studied the decay chain of the moduli with the package PYTHIA~\cite{Sjostrand:2006za} for a few sample benchmark models in the G2-MSSM, 
assuming a stable Wino-like chargino, and found energy distributions for the Wino-like neutralinos and charginos which are typically peaked at 
$E/m_\chi \sim 10$ and with very broad tails up to the kinematical threshold; among decay products, the number of charginos is typically about 3 times larger 
than the number of neutralinos.   

The injected ultra-relativistic particles lose energy via scattering on thermal bath states. Were these processes inefficient, the non-thermal DM generation
would give rise to a model of the Universe with warm or even hot DM, a possibility which has been investigated, e.g., in Refs.~\cite{Hisano:2000dz,Gelmini:2006vn}.
As a first rule of thumb, the energy depletion is efficient whenever the relative energy loss rate times the time interval the over which the process
is active, which we indicate as $\Delta \tau$, is larger than 1:
\begin{equation}
\label{eq:thc}
 \left(-\frac{1}{E}\frac{dE}{dt}\right)\cdot \Delta \tau >1\,.
\end{equation}
In our case this condition needs to hold from the relativistic regime down to the non-relativistic low-temperature environments induced by the reheating phase. 
The expression for $-dE/dt$ is in the form:
\begin{equation}
\label{elos}
  -\frac{dE}{dt}=\int d{E}^{\prime} \left(E-{E}^{\prime}\right)\frac{d\Gamma}{d{E}^{\prime}}\left(E,{E}^{\prime}\right)
\end{equation}
where ${\Gamma}(E)$ the scattering rate for the process under scrutiny, integrated over the phase space distribution functions of the thermal bath particles 
in the initial state and the phase space of the out-scattered particles. The expressions we will report below are derived in the limit of small momentum transfer 
between the non-thermally produced states and the thermal bath particles; the latter on average have energies equal to about $3\,T$. The small 
momentum transfer approximation holds whenever the non-thermal particles are non-relativistic in the CM frame of the scattering process, namely 
for~\cite{Hisano:2000dz}:
\begin{equation}
\label{wdmhisano}
  m_\chi^2 \gtrsim 6 T E\,.
\end{equation}
Assuming instantaneous production at reheating, this relation can be translated into:
\begin{equation}
 T_{RH}  \lesssim 1.7 \,\mbox{GeV}\; \left(\frac{m_\chi}{100 \,\mbox{GeV}}\right)\left(\frac{10}{E/m_\chi}\right)\,,
\end{equation}
a condition which is satisfied in the region of the G2-MSSM parameter space providing a viable DM candidate.

Charginos lose energy via electromagnetic interactions and their energy loss rate takes the form \cite{Reno:1987qw,Braaten:1991jj}: 
\begin{equation}
  \left(-\frac{dE}{dt}\right)_{\chi^\pm}=\frac{\pi {\alpha}^{2} {T}^{2}}{3}\Lambda
\end{equation}
with $\Lambda \sim O(1)$. The elastic scattering of a Wino-like neutralino on a background lepton is very inefficient, since it proceeds
via a $Z$ boson or a slepton exchange and the corresponding amplitudes are suppressed, respectively, by the tiny higgsino fraction in the LSP
and by the slepton masses, which in the G2-MSSM are very heavy. Whenever kinematically allowed, the dominant effect is the inelastic
scattering into the charged Wino, which is mediated by a $W$ boson. There are then two effects making a neutralino produced in the decay of 
the modulus lose energy, namely the energy loss in the inelastic scattering itself and the fact that the produced chargino will efficiently lose energy. 
For relativistic neutralinos, the inelastic scattering rate and the energy loss rate in inelastic scatterings are, respectively, given by:
\begin{eqnarray}
  \Gamma_{\chi^0\rightarrow \chi^\pm}
  &=& \sum_{(a,b)} \frac{2 {\tilde{g}}_{Wab} \, G_{\rm F}^2}{{\pi}^3 } \exp{\left(-\frac{m_\chi \Delta m_\chi}{2ET}\right)}  \; T^4 \frac{E}{m_\chi}\left(6\frac{ET}{m_\chi}+\Delta m_\chi \right) \\
  \left(-\frac{dE}{dt}\right)_{\chi^0\rightarrow \chi^\pm}
  &=& \sum_{(a,b)} \frac{16 {\tilde{g}}_{Wab} \, G_{\rm F}^2}{\pi^3 } \exp{\left(-\frac{m_\chi \Delta m_\chi}{2ET}\right)} \;T^5 \left(\frac{E}{m_\chi} \right)^3 \left(8\frac{ET}{m_\chi}+\Delta m_\chi\right)\,.
\end{eqnarray}
where, considering the generic process in which the heat bath particle $a$ is scattered into the particle $b$ via an interaction vertex with a W boson, we 
have included in the coefficient ${\tilde{g}}_{Wab}$ the product of the number of internal degrees for $a$, that for $b$, as well as a rescaling factor in case
the coupling constant in the vertex is different from the SU(2) gauge coupling $g$ (e.g., for the scattering process 
${\chi}^{0}+{e}^{\pm} \rightarrow {\chi}^{\pm}+{\nu}_{e}$, 
 ${\tilde{g}}_{Wab} = 8$); the sum goes over any $(a,b)$ thermal bath particle pairs.

\begin{figure}[htbp]
\centering
\subfigure[\hspace{0.1cm} Case 1:${m}_{\chi}=103.5$ GeV, $T_R=100$ MeV, $E/m_\chi=1.005$.]{\includegraphics[width=8 cm, height= 6 cm, angle=360]{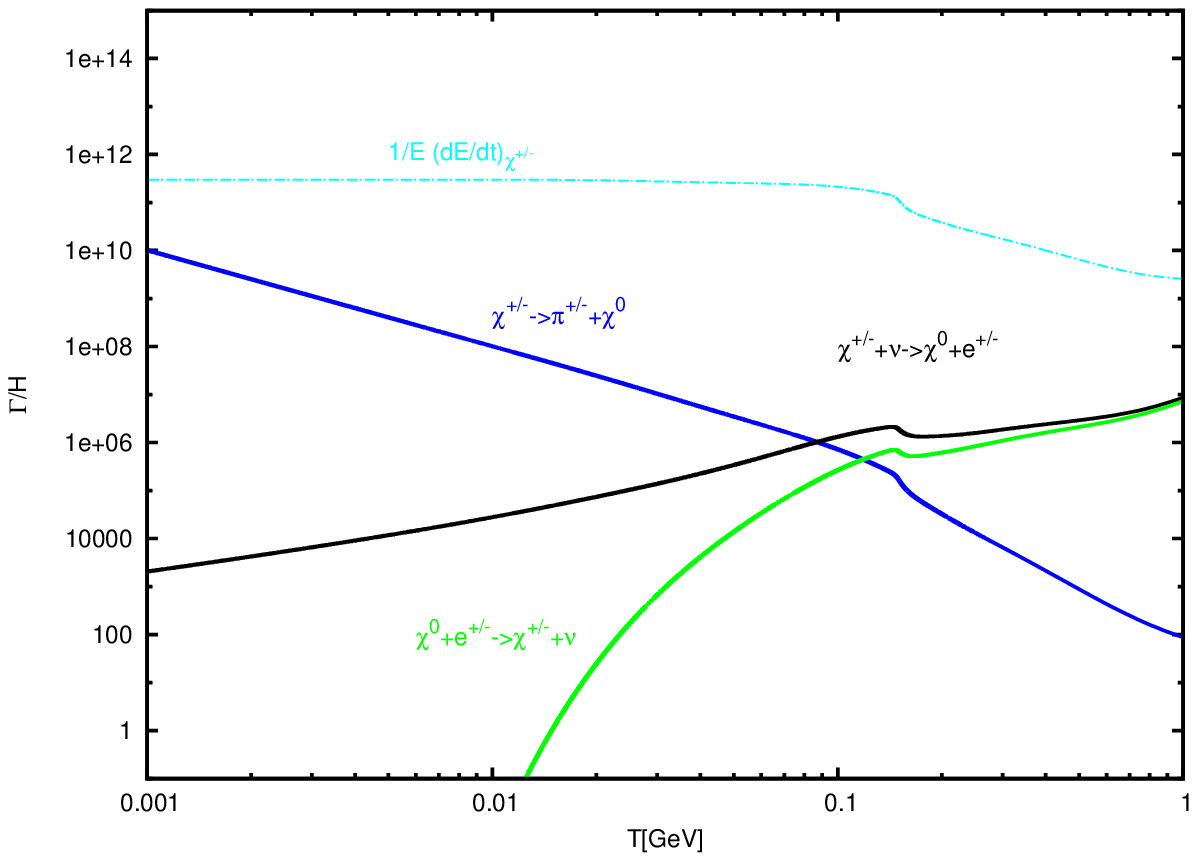}}
\subfigure[\hspace{0.1cm} Case 2:${m}_{\chi}=103.5$ GeV, $T_R=100$ MeV, $E/m_\chi=10$.]{\includegraphics[width=8 cm, height= 6 cm, angle=360]{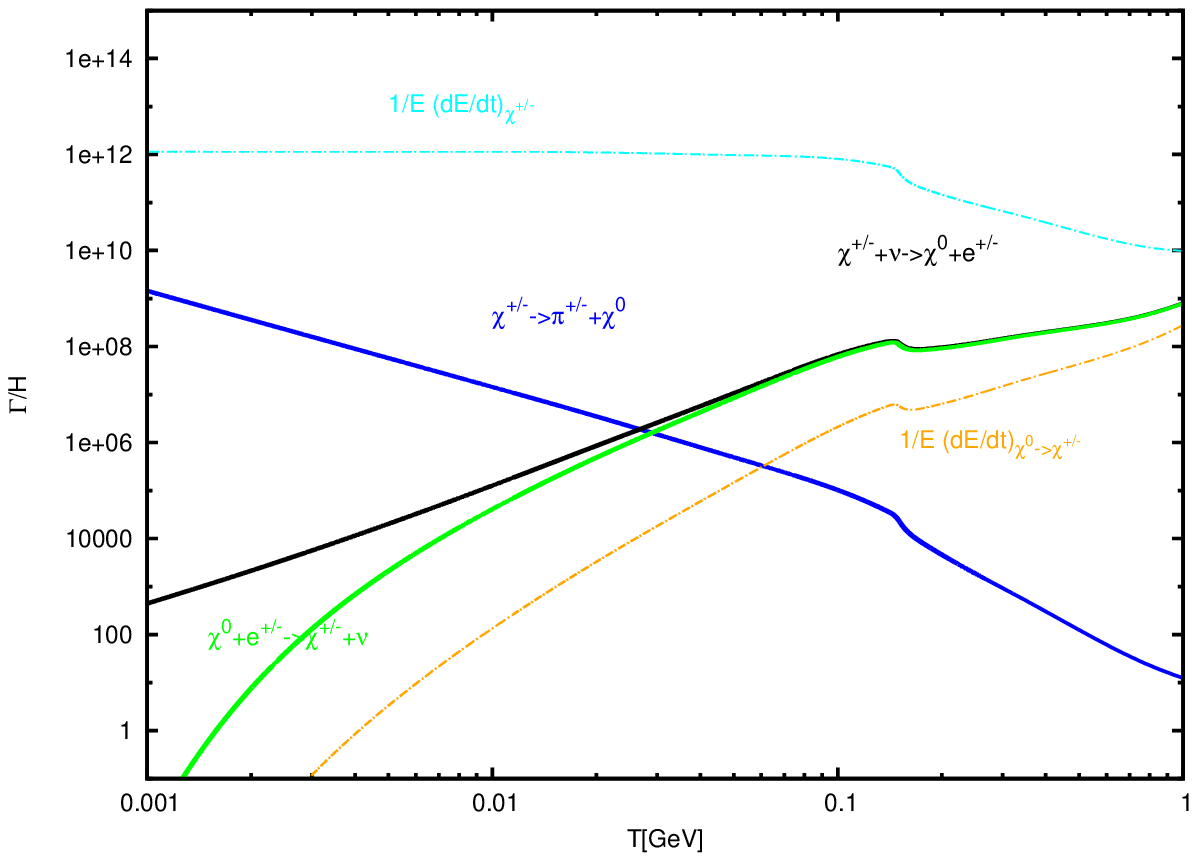}}\\  
\subfigure[\hspace{0.1cm} Case 3:${m}_{\chi}=300$ GeV, $T_R=900$ MeV, $E/m_\chi=1.005$.]
    {\includegraphics[width=8 cm, height= 6 cm, angle=360]{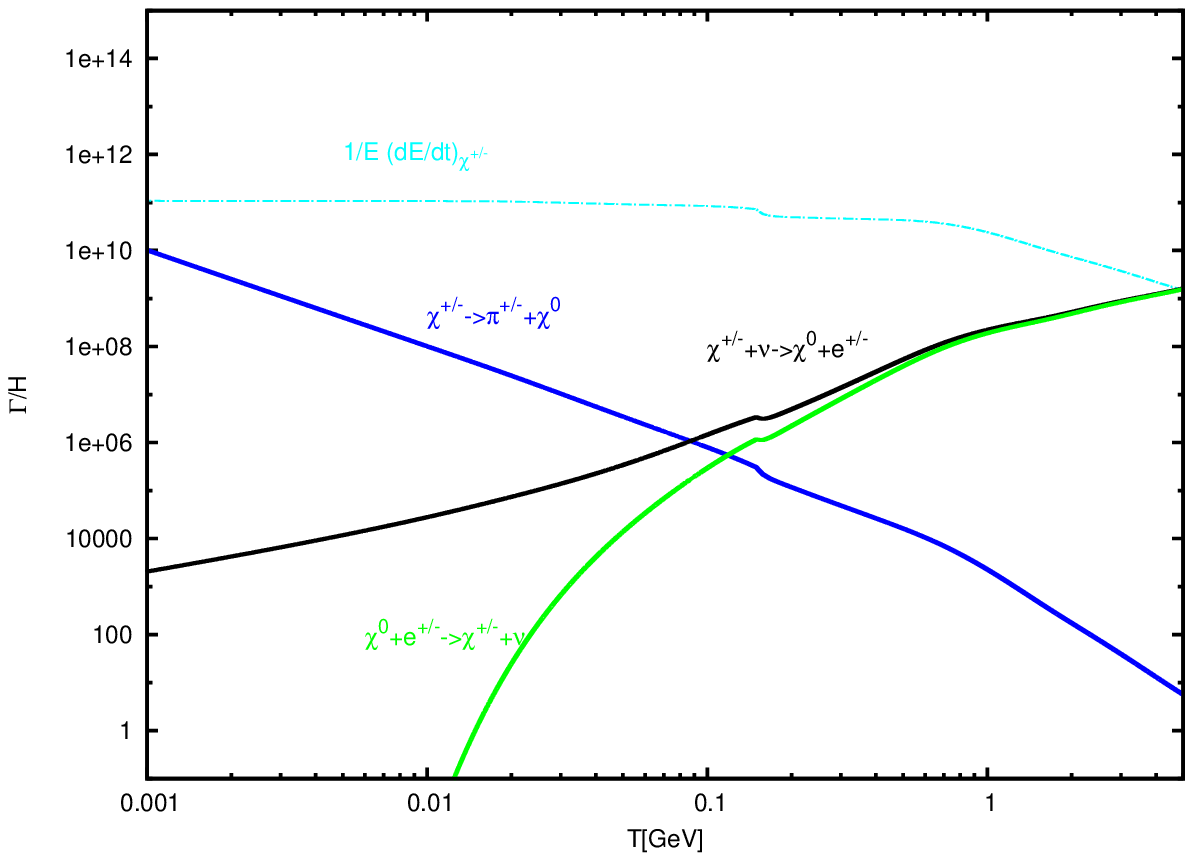}}
\subfigure[\hspace{0.1cm} Case 4:${m}_{\chi}=300$ GeV, $T_R=900$ MeV, $E/m_\chi=10$.]
    {\includegraphics[width=8 cm, height= 6 cm, angle=360]{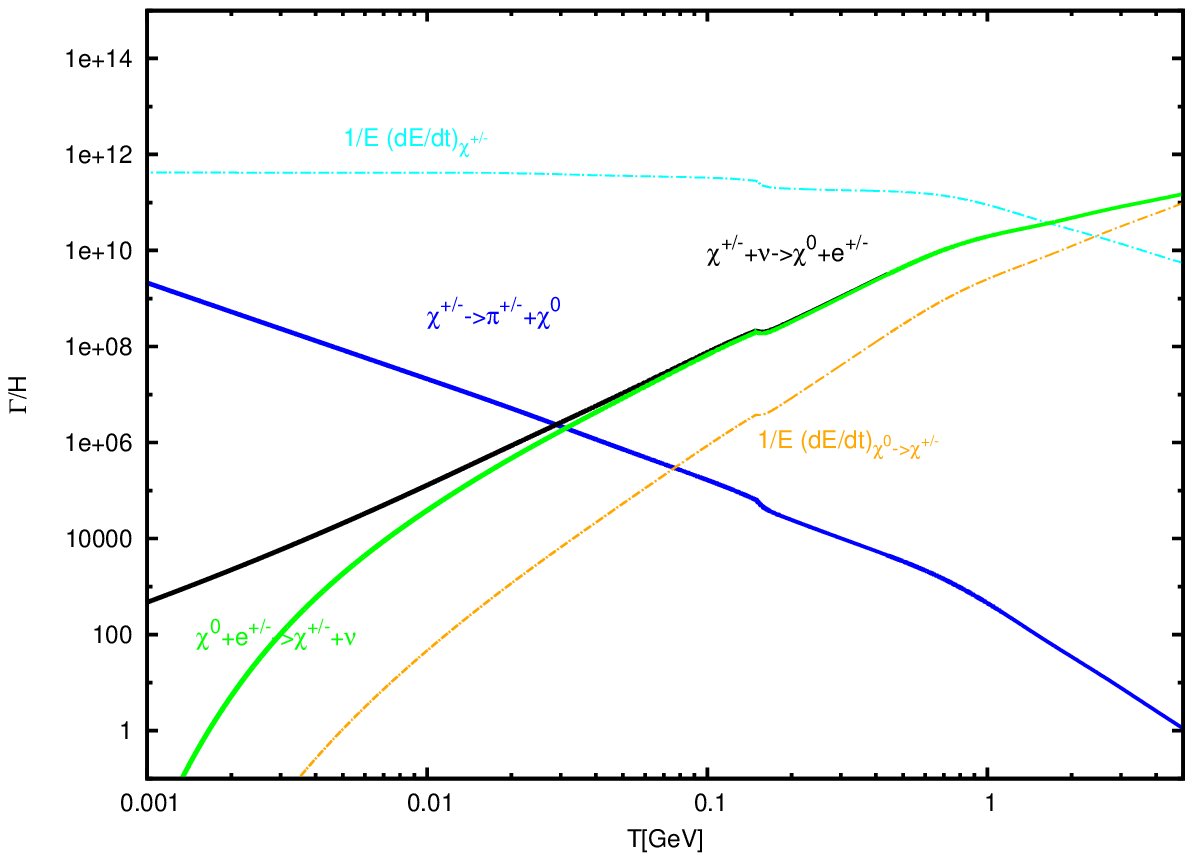}}
\caption{Ratios between the effective rate of energy loss rate $-1/E \cdot dE/dt$ (dashed lines), or of the scattering/decay rate $\Gamma$ (solid lines), to the 
Universe expansion rate $H$, for a few processes involving charginos and neutralinos. The upper panels refer to a G2-MSSM DM model with 
$m_\chi=103.5$~GeV and  ${T}_{RH}=100$~MeV, while the lower panels to one with $m_\chi=300$~GeV and  ${T}_{RH}=900$~MeV; the plots on the
left hand-side refer to non-relativistic particles, $E/m_\chi = 1.005$, while those on the right-hand side correspond to a sample relativistic case, $E/m_\chi =10$.    
}
\label{fig:energyloss}
\end{figure}

In Fig.~\ref{fig:energyloss} we consider two of the G2-MSSM singled out in the previous Section as models embedding a viable DM candidate, at the 
light and heavy ends of the mass range displayed in Fig.~\ref{fig:fig5}, i.e. two models with Wino masses, respectively, of 103.5 and 300  GeV, obtained 
for $D_{X_i}=16$, $\delta = -3.5$ and $\delta=-3$ and gravitino masses of 107 and 460~TeV, and corresponding to scenarios with approximate reheating temperatures of 100 MeV and 900 MeV. For such models  we plot ratios of scattering and decay rates $\Gamma$, or of relative energy loss rates $-1/E \cdot dE/dt$, to the Universe expansion rate $H$; in the panels on the right-hand side, results are shown for relativistic particles, $E/m_\chi =10$, while 
on the left-hand side the non-relativistic limit is considered, $E/m_\chi = 1.005$. To sketch the efficiency of the chargino energy losses, the 
appropriate timescale $\Delta \tau$ in Eq.~(\ref{eq:thc}) is the shortest between the chargino lifetime and the timescale for back-scattering of the chargino into the neutralino, i.e. the rule of thumb condition in Eq.~(\ref{eq:thc}) holds whenever 
the curves in plots corresponding to the chargino energy loss lie above the curves for the decay rate and the inelastic scattering rate. More quantitatively, for 
the two processes, these ratios  are:
\begin{eqnarray}
  \left(\frac{\Delta E}{E}\right)_{\chi^\pm, {\rm 1-life}} &\approx& 
  1.86 \cdot 10^2 \Lambda \left(1-\frac{m_\pi^2}{\Delta m_{\chi}^2}\right)^{-1/2} \left(\frac{T}{1 \,\mbox{MeV}}\right)^2
  \left(\frac{160 \,\mbox{MeV}}{\Delta m_\chi}\right)^3 \left(\frac{100 \,\mbox{GeV}}{m_\chi}\right) \\
  \left(\frac{\Delta E}{E}\right)_{\chi^\pm, {\rm 1-scat.}} &\approx& 
  6.77 \cdot 10^{-2}  \Lambda \left(\frac{1 \, \mbox{GeV}}{T}\right)^3 \left(\frac{m_\chi}{100 \,\mbox{GeV}}\right)
  \left(\frac{10}{E/m_\chi}\right)^3 \,,
 \end{eqnarray}
with the smallest of these being much larger than one in all cases except for relativistic charginos injected at temperatures of the order of 1~GeV or larger. The latter 
is however the regime in which inelastic scatterings turning a neutralino into a chargino and viceversa are extremely efficient (as shown in the plots the rate for this
process is many orders of magnitude larger than $H$) and the energy loss rate via this process is also very large (the relevant timescale is now $\sim H^{-1}$). 
This shows that the relativistic charginos injected at any of the temperatures of interest in our model instantaneously thermalize. For what regards neutralinos, 
in the relativistic limit, the energy depletion is guaranteed by inelastic scatterings and by chargino energy losses down to background temperatures of about 
2~MeV; however when becoming non-relativistic and at low temperatures, the rate for inelastic scatterings becomes smaller than $H$ and the assumption
of kinetic equilibrium may not hold any more. 
To study the evolution of the system at low temperature and model kinetic decoupling, we follow the approach of Bringmann and 
Hofmann~\cite{Bringmann:2006mu} (see also \cite{Bringmann:2009vf}) who have developed a formalism to treat kinetic decoupling starting from the 
Boltzmann equation for the phase-space distribution function of the WIMP DM candidate; we extend here their treatment to the case of two co-annihilating 
particles. Let $f_{\chi^0}(p,t)$ and $f_{\chi^\pm}(p, t)$ be the phase space distribution functions for, respectively, the neutral and charged Winos. 
We have just shown that charginos are kept into kinetic equilibrium at all temperatures of interest for your problem, so we can assume that the shape of 
the chargino distribution traces the thermal distribution function, namely:
\begin{equation}
  f_{\chi^\pm}(p, t) \propto f_{\chi^\pm}^{eq}(p, t) \,.
\end{equation}
On the other hand, the distribution function of the neutralinos could have a shape which is slightly different from the thermal one, 
since we have shown that energy losses may not be efficient in the non-relativistic regime; this departure is parametrized defining the temperature of 
neutralinos $T_{\chi^0}$ through the second moment of the distribution function:
\begin{equation}
 \int \frac{d^3p}{(2\pi)^3}\,g_{{\chi}^{0}} \,p^2 \,f_{\chi^0}(p,t) \equiv 3 m_\chi \, T_{\chi^0}(t) \,n_{\chi^0}(t) \,.
\end{equation} 
For neutralinos in kinetic equilibrium, $T_{\chi^0}$ coincides with the thermal bath temperature; after kinetic decoupling the neutralino temperature 
will scale instead as $T_{\chi^0} \propto T^2$. 

The two distribution functions obey the system of coupled Boltzmann equations:
\begin{eqnarray}
  \label{eq:psds} 
  \left(\partial_t-H\mathbf{p}\cdot \nabla_{\mathbf{p}}\right)f_{\chi^0}(p,t) & = &\frac{1}{E} \,{\hat{\mathbf C}}_{\chi^0}[f_{\chi^0},f_{\chi^\pm}] \\ \nonumber
  \left(\partial_t-H\mathbf{p}\cdot \nabla_{\mathbf{p}}\right)f_{\chi^\pm}(p,t) & = &\frac{1}{E} \,{\hat{\mathbf C}}_{\chi^\pm}[f_{\chi^0},f_{\chi^\pm}]\,,
\end{eqnarray}
where ${\hat{\mathbf C}}$ stands for the collisional operator, embedding all interactions involving neutralinos and charginos, namely
annihilation and scattering processes, as well as the production of neutralinos and chargino from moduli decays and the neutralino source from chargino 
decays. Integrating these equation over phase space one obtains two equations for the time evolution of the neutralino and chargino number densities:
\begin{eqnarray}
   \label{eq:nds}
   \frac{dn_{\chi^0}}{dt} + 3\,H\,n_{\chi^0} &=&  
    \left(  \widetilde{\Gamma}_{\chi^0 \leftrightarrow \chi^\pm} +\Gamma_{\chi^\pm}\right) \left[g_{\chi^0} n_{\chi^\pm} -{g}_{\chi^\pm} n_{\chi^0} \exp{\left(-\frac{\Delta m_\chi}{T}\right)}\right] 
      - \langle\sigma v\rangle_{\chi^0 \chi^0} \left[n_{\chi^0}^2 - (n_{\chi^0}^{eq})^2 \right] \\ \nonumber
   &&-  \langle\sigma v\rangle_{\chi^0 \chi^\pm} \left[n_{\chi^0} n_{\chi^\pm} - n_{\chi^0}^{eq} n_{\chi^\pm}^{eq} \right]
\\ \nonumber\
   \frac{dn_{\chi^\pm}}{dt} + 3\,H\,n_{\chi^\pm} &=&
    \left( \widetilde{\Gamma}_{\chi^0 \leftrightarrow \chi^\pm} + \Gamma_{\chi^\pm} \right) \left[{g}_{\chi^\pm} n_{\chi^0} \exp{\left(-\frac{\Delta m_\chi}{T}\right)} - g_{\chi^0} n_{\chi^\pm}\right]
     - \langle\sigma v\rangle_{\chi^\pm \chi^\pm} \left[n_{\chi^\pm}^2 - (n_{\chi^\pm}^{eq})^2 \right] \\ \nonumber
   &&-  \langle\sigma v\rangle_{\chi^\pm \chi^0} \left[n_{\chi^\pm} n_{\chi^0} - n_{\chi^\pm}^{eq} n_{\chi^0}^{eq} \right]
     + \sum_i \frac{B_{X_i}}{m_{X_i}} \Gamma_{X_i} \rho_{X_i} \,. 
\label{eq:singlen}
\end{eqnarray}
In these equations, the first term on the right hand sides accounts for inelastic scatterings of neutralinos into charginos and decays of charginos
into neutralinos (as well as the inverse processes); $g_{\chi^0}$ and $g_{\chi^\pm}$ are the number of internal degrees of freedom for the neutralino and chargino,
while $\Gamma_{\chi^\pm}$ is the chargino decay rate as obtained including the two contributions in Eq.~(\ref{eq:chdecay}).
For inelastic scatterings we have assumed that: {\sl i)} the diagram with $W$ boson exchange in the t-channel is the dominant one (since those with sfermion exchanged are highly suppressed); {\sl ii)} the momentum transfer in the t-channel is small and the collision term can be computed expanding in 
its powers, see also~\cite{Bringmann:2006mu,Bringmann:2009vf}; {\sl iii)} $\Delta m_\chi$ and $T$ are small with respect to $m_\chi$ and only the lowest
order terms in an expansion in $\Delta m_\chi$ and $T$ give sizable contributions; under these assumptions, we find:
\begin{equation}
   \widetilde{\Gamma}_{\chi^0 \leftrightarrow \chi^\pm} =\sum_{(a,b)} \frac{{\tilde{g}}_{Wab}8G_{\rm F}^2}{{\pi}^3} T^3 \left(\Delta m_\chi^2 + 6 \Delta m_\chi T+12 T^2 \right) \,,
\end{equation}
(some further details and a sketch of the derivation of this expression is given in Appendix~(\ref{appendix1})). When including pair annihilation terms in 
Eq.~(\ref{eq:nds}) we have taken advantage of the fact that it involves non relativistic particles annihilating mainly via S-wave processes and 
hence the cross section has a very mild dependence on momentum, allowing then us to write an expression analogue to thermal case also when the neutralino distribution function starts deviating from the kinetic equilibrium value. Furthermore, since the relativistic particles injected from the moduli decays mostly lose energy as charginos, decaying afterwards into neutralinos, we have simplified the treatment including these as a source function of "thermal" charginos only. Obviously, summing the two equations one retrieves Eq.~(\ref{eq:dm}) with $n_\chi$ being the sum of the number density for the two coannihilating species.

Taking the second moment of the first equation in the system in Eq.~(\ref{eq:psds}), one finds that the neutralino temperature $T_{\chi^0}$ obeys the relation:
\begin{equation}
   \label{eq:temp}
   \frac{dT_{\chi^0}}{dt} + 2 H  T_{\chi^0} = \left[ \left( \widetilde{\Gamma}_{\chi^0 \leftrightarrow \chi^\pm}   + \Gamma_{\chi^\pm} \right) g_{\chi^0} \frac{n_{\chi^\pm}}{n_{\chi^0}} \right] \,(T- T_{\chi^0})
\end{equation}
(the derivation of this equation is also sketched in the appendix). 

The numerical solution of the problem proceeds now analogously to what done so far. After the appropriate change of variables, the system in Eq.~(\ref{eq:nds}) 
replaces Eq.~(\ref{eq:dm}) in the system of Eq.~(\ref{eq:system}). The explicit solution for $n_{\chi^0}(t)$ and $n_{\chi^\pm}(t)$ are then implemented 
in Eq.~(\ref{eq:temp}) to find $T_{\chi^0}(t)$.

\begin{figure}[htbp]
 \begin{minipage}[htb]{8cm}
   \includegraphics[width=8 cm, height= 6 cm, angle=360]{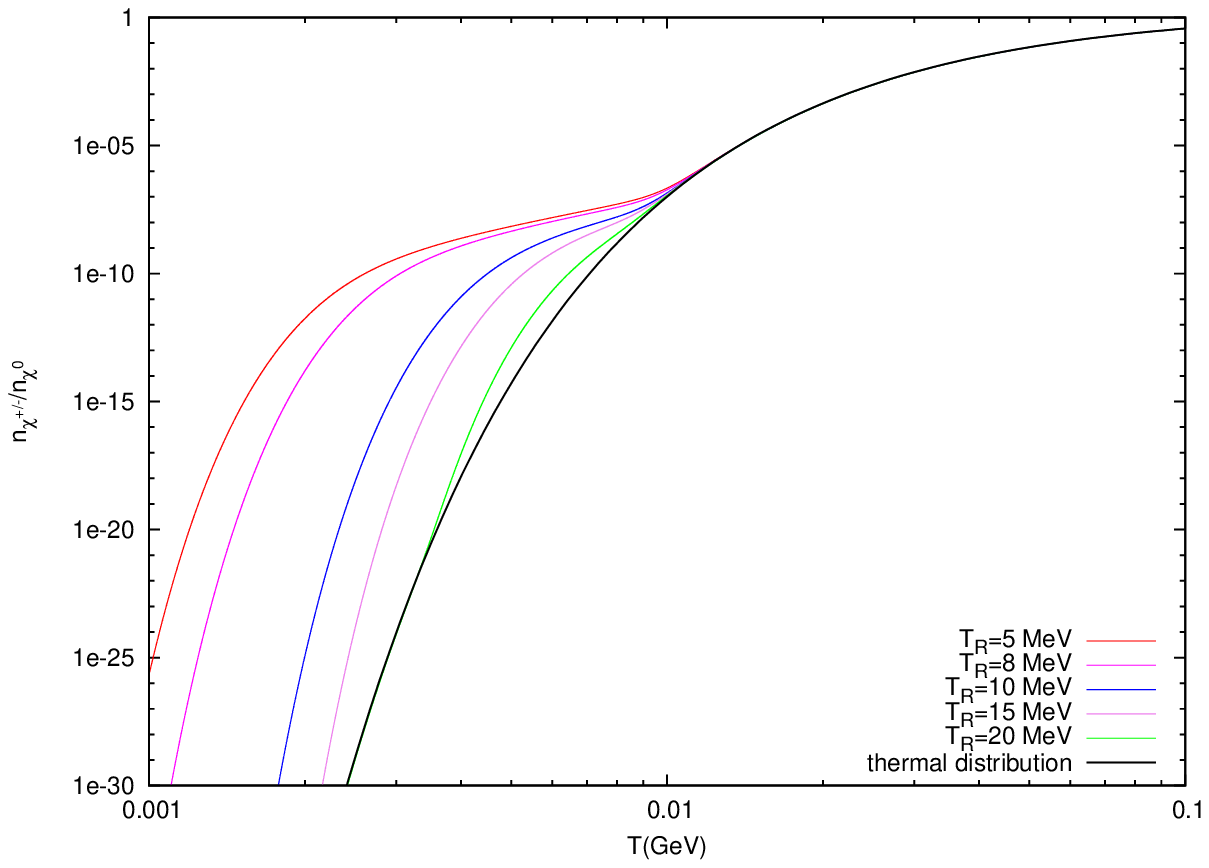}
 \end{minipage}
 \ \hspace{3mm} \
 \begin{minipage}[htb]{8cm}
   \includegraphics[width=8cm, height= 6 cm, angle=360]{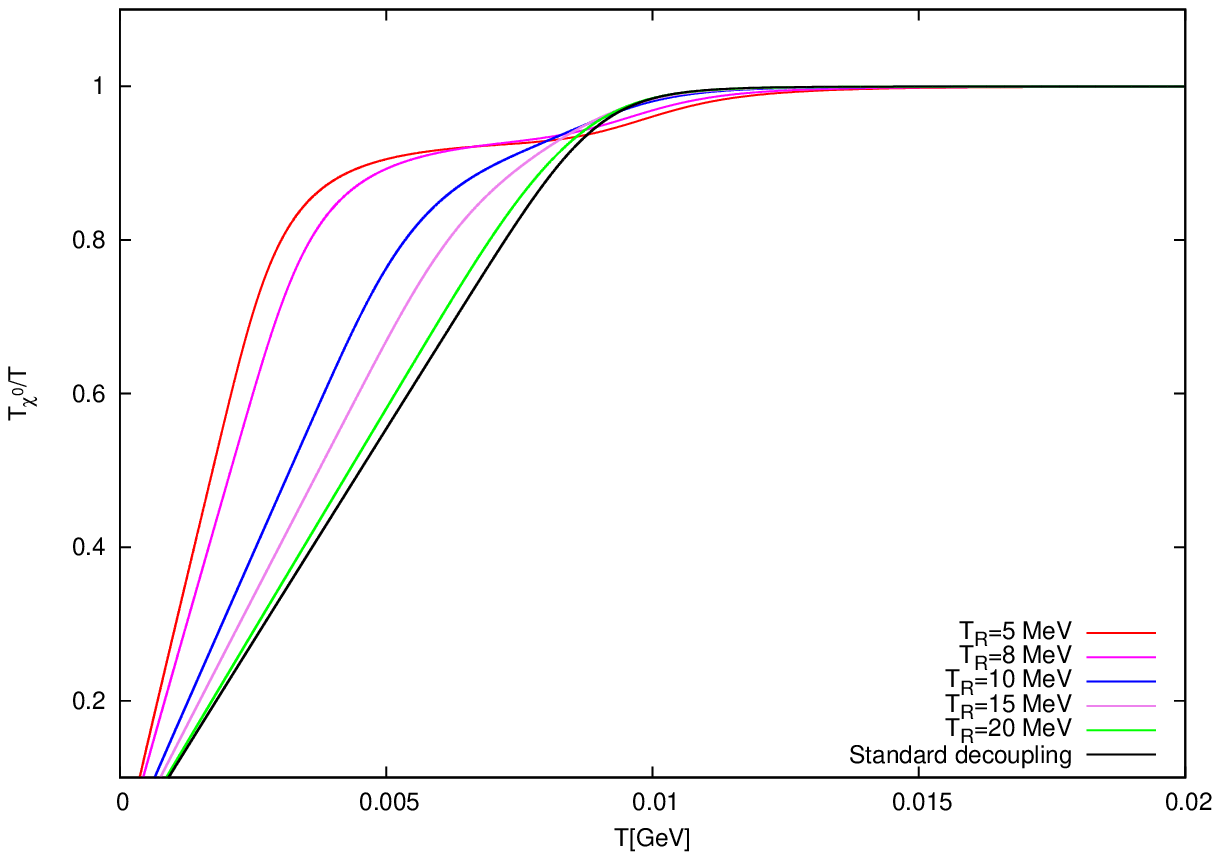}
 \end{minipage}
\caption{Left panel: ratio of the chargino number density over the neutralino number density for several values of ${T}_{RH}$. 
Right panel: Ratio $T_{{\chi}_{0}}/T$ as function of
 the temperature of the thermal bath for same values of $T_{RH}$. Plots are obtained for a Wino with mass equal to 200~GeV, however results depend
 only on the chargino-neutralino mass splitting which is about 160~MeV in the scenario under consideration.}
\label{fig:Tsever}
\end{figure}

Our first application is to the G2-MSSM models singled out in the previous Section as cosmologically favored. As we had guessed in the analysis we performed 
at the level of energy loss and scattering rates and shown graphically in Fig.~\ref{fig:energyloss}, the departure from kinetic equilibrium tends to be at a 
temperature sensibly lower than the nominal reheating temperatures for these models (which are of the order of 100~MeV or larger). The numerical solution
indeed shows that the ratio $n_{\chi^\pm}/n_{\chi^0}$ tends to follow very closely the ratio of the thermal equilibrium number densities 
$n_{\chi^\pm}^{eq}/n_{\chi^0}^{eq}$ over the whole phase of DM production in the moduli decays, as well as at later times. The solution of the equation for the
neutralino temperature shows that kinetic equilibrium is maintained up to a temperature of the order of 10~MeV, independently of the neutralino  mass since, 
in the non-relativistic limit, the inelastic scattering rate (which together with chargino electromagnetic interactions enforces the equilibrium) depends only
on the chargino-neutralino mass splitting which is essentially the same over the whole range of selected models.
The transition between $T_{{\chi}_{0}}=T$ to the regime $T_{{\chi}^{0}} \propto T^2$ takes place on relatively short timescales; since at 10~MeV non-thermal 
production has become irrelevant, we would have found the very same scaling when computing the kinetic decoupling for a population of thermal particles:
the evolution of the number density ratio and of $T_{{\chi}^{0}}$ for G2-MSSM DM models are those shown as black lines in Fig.~\ref{fig:Tsever} and 
labelled, respectively, 'thermal distributions' and 'standard decoupling'.

To illustrate the impact of non-thermal production and non-standard cosmologies on the kinetic decoupling process, we allow then for a slight variant to 
the underlying particle physics framework, still referring to a pure Wino as DM candidate but assuming now that the reheating temperature can be reduced to
values much closer to the bound from BBN than in the G2-MSSM. In Fig.~\ref{fig:Tsever} one sees a modification with respect to the standard case when 
the gap between reheating temperature and standard kinetic decoupling temperature is reduced, i.e. for $T_{\rm RH}$ equal to about 20~MeV or lower: 
The additional DM source makes the ratio $n_{\chi^\pm}/n_{\chi^0}$ differ from the ratio of thermal distributions. The impact on $T_{{\chi}_{0}}$ is two folded:
the chargino decays tend to populate the system with neutralinos that are on average more energetic than for a thermal distribution, delaying the onset of the
regime $T_{{\chi}^{0}} \propto T^2$ and making the transition into this regime to be less sharp; at the same time, if $T_{\rm RH}$ is so low that reheating increases
significantly the expansion rate of the Universe $H$ at the time of kinetic decoupling ($T_{\rm RH} = 8$ and 5~MeV in the plot) the departure from
$T_{{\chi}_{0}}=T$ tends to be anticipated. This latter feature was already pointed in \cite{Gelmini:2008sh}, showing that the non-thermal production   
could induce higher kinetic decoupling temperatures compared to the standard case; in case of Wino DM, however, the production and decay or charginos 
in the moduli decay has always a larger impact. 
The kinetic decoupling temperature is directly related to the minimum mass scale for structures in the Universe; we have shown here that even in case 
of injection of particles with efficient energy losses on the thermal bath, the low-temperature non-thermal production can leave an imprint on structure 
formation. The development of a precise numerical treatment of the kinetic decoupling is then a valuable tool to test this class of models.  

Finally, in the examples considered here, we find a marginal change in the DM relic density when computing it in the case when we trace the
the number densities of the individual coannihilating species as opposed to the case when a single equation for the sum of number densities
is solved; this is due to the fact that the departure of the ratio $n_{\chi^\pm}/n_{\chi^0}$ from the ratio of thermal distributions
takes place only when such quantity is very small (moreover, in our examples, the annihilation rates for each of the coannihilation channels  are
comparable). Considering however models for which crossing symmetry arguments between annihilation and scattering cross sections are even more
severely violated, one should find cases in which the standard thermal assumption is invalid at higher temperatures, 
possibly even close to the chemical freeze out temperature; in those cases there should be a sizable change in the relic abundance as well and
the formalism we developed would be suitable for an accurate computation of the relic density for such case.

\section{Conclusions}

Non-thermal dark matter production is a viable alternative to the standard paradigm of WIMPs as thermal relics. It is a well-motivated scenario arising 
naturally in several particle physics frameworks, including SUSY standard model extensions within supergravity and superstring theories.
Moreover, an epoch of entropy injection at an intermediate phase between the reheating at the end of inflation and the onset of BBN dilutes 
dangerous long-lived relics, such as gravitinos. We have reviewed how to introduce a system of differential equations to treat a generic case 
of non-thermal dark matter generation and implemented an efficient and accurate numerical scheme for the computation;  such scheme has been 
interfaced to \ds\ numerical package and will be released together with an upcoming version of the code. The interest in this scenario has been 
recently boosted by the fact that, generically, it favors dark matter candidates with pair annihilation cross sections larger than in the thermal
WIMP framework, possibly suggesting a connection with the very large annihilation rates which would be needed to explain with a dark matter
induced component recently detected anomalies in cosmic-ray lepton fluxes, such as the rise in the positron fraction measured by the PAMELA 
detector; we have considered, within a toy model, what range of reheating temperatures would follow from such identification. Focussing
on SUSY models, we have discussed the impact of this non-standard cosmology in selecting the preferred mass scale for the lightest SUSY 
particle as dark matter candidate, an issue with a direct impact, e.g., on the interpretation of new physics eventually discovered at accelerators 
including the LHC. In the second part of the paper we have concentrated on a more predictive model, the G2-MSSM, and questioned in further
details the underlying assumptions in the standard solution of the Boltzmann equation for the dark matter component; in particular, we discussed 
how to verify whether kinetic equilibrium holds along the whole phase of dark matter generation, as well as the validity of  the factorization 
usually implemented to rewrite a system of coupled Boltzmann equation for each number density of a set coannihilating particles as a single 
equation for the sum of all the number densities. As a byproduct we developed here a formalism to compute the kinetic decoupling temperature 
for a system of coannihilating species, which can be applied also to other particle physics frameworks, also in case standard thermal relics
within a standard cosmology.

\acknowledgments

We would like to thank Bobby Acharya for discussions on the G2-MSSM, Yann Mambrini, Umberto De Sanctis and Marco Nardecchia comments
and suggestions, and Sfefano Profumo for providing us fits to the Pamela and Fermi lepton excess reproduced in Fig.~\ref{fig:fig3}. 
We also would like to thank the Galileo Galilei Institute for  Theoretical Physics for hospitality during part of the time this work  
was developed.

\appendix

\section{Evaluation of the collisional operators}
\label{appendix1}

In this appendix we will sketch how to compute the operators $\mathbf{\hat{C}}$ as introduced in the system of coupled Boltzmann equations (\ref{eq:psds}). As a 
 sample term we discuss how to deal with the contribution to the neutralino collisional operator coming from inelastic processes of the type:
\begin{equation}
 {\chi}^{0}(P)+a(K) \leftrightarrow {\chi}^{\pm}\left(P^{\prime}\right)+b({K}^{\prime}) 
\end{equation}
where $a$ and $b$ are thermal background particles, with four-momenta, respectively, ${K}\equiv \left(k,\mathbf{k}\right)$ and 
${K}^{\prime} \equiv \left({k}^{\prime},{\mathbf{k}}^{\prime}\right)$, while 
$P \equiv \left(E,\mathbf{p}\right)$ and ${P}^{\prime} \equiv \left({E}^{\prime},{\mathbf{p}}^{\prime}\right)$ denote the four-momenta
of neutralino and chargino. Summing over all avalaible thermal bath pairs $(a,b)$, 
such contribution to the collisional operator, normalized to the neutralino energy $E$, takes the form:
\begin{eqnarray}
 \frac{{\hat{\mathbf{C}}}_{{\chi}^{0},\rm is}}{E}[{f}_{{\chi}^{0}},{f}_{{\chi}^{\pm}}] &=& \sum_{(a,b)} {\tilde{{g}}_{Wab}} {g}_{{\chi}^{\pm}} 
\int \frac{{d}^{3}k}{{\left(2\pi\right)}^{3}2k}\int \frac{{d}^{3}{k}^{\prime}}{{\left(2\pi\right)}^{3}2{k}^{\prime}}
\int \frac{{d}^{3}{p}^{\prime}}{{\left(2\pi\right)}^{3}2{E}^{\prime}}\frac{{|\bar{M}|}^{2}_{ab}}{2E}
{\left(2\pi\right)}^{4}\delta^4({P}^{\prime}+{K}^{\prime}-P-K) \cdot \nonumber\\
&&\cdot \left[f_b({k}^{\prime})(1-f_a(k)){f}_{{\chi}_{\pm}}({p}^{\prime})
-f_a(k)(1-f_b({k}^{\prime})){f}_{{\chi}^{0}}(p)\right]\,;
\label{eq:a2}
\end{eqnarray}
where $|\bar{M}|^{2}_{ab}$ is the modulus squared of the scattering amplitude, averaged over the initial spin states and summed over the final 
spin states. In the following we assume that $a$ and $b$ are massless and described by Fermi-Dirac distribution functions  $f_a(k,t)$ and 
$f_b({k}^{\prime},t)$ (to shorten the notation the indices $a$ and $b$ will be dropped). The exchanged four-momentum is indicated as
$\left(\omega,\mathbf{q}\right) \equiv \left(E-{E}^{\prime},\mathbf{p}-{\mathbf{p}}^{\prime}\right)$; for kinematical reason, the transferred momentum 
is constrained to be of the order of the heat bath temperature and is small compared to the initial energy and mass of the neutralino. The ratio 
between the mass splitting chargino-neutralino and the neutralino mass will be also assumed as a small parameter.  Furthermore we have initial 
conditions such that neutralinos have nearly thermal distributions, implying $v \sim \sqrt{T/m\chi} \ll 1$, being the 
neutralino velocity defined as $\mathbf{v} \equiv \mathbf{p}/E$. Under such conditions, we can just take the non-relativistic limit of the collision 
term and, in addition, expand it respect to the quantities $T/{m}_{\chi}$ and ${\Delta m}_{\chi}/{m}_{\chi}$.
Using these informations we can eliminate the dipendece on ${\mathbf{p}}^{\prime}$ by 
Taylor expanding $f_{{\chi}^{\pm}}(p^{\prime})$ as:
\begin{equation}
 f_{{\chi}^{\pm}}({p}^{\prime})\simeq f_{{\chi}^{\pm}}(p)-\mathbf{q} \cdot {\nabla}_{\mathbf{p}}f_{{\chi}^{\pm}}(p)
+\frac{1}{2}{\left(\mathbf{q} \cdot {\nabla}_{\mathbf{p}}\right)}^{2}f_{{\chi}^{\pm}}(p)+... 
\end{equation} 
This allows us to freely integrate over the three-momentum component of the delta function in Eq.~(\ref{eq:a2}). Using now the relation:
\begin{equation}
 f(k^{\prime})(1-f(k))=\exp\left(-\frac{\omega}{T}\right) f(k)(1-f(k^{\prime}))\simeq \exp\left(-\frac{\omega}{T}\right) f(k)
\end{equation}
with:
\begin{equation}
  \exp\left(-\frac{\omega}{T}\right)\simeq \exp\left(\frac{{\Delta m}_{\chi}}{T}\right)\left(1-\frac{{\Delta m}_{\chi}v^2}{2T}-
  \frac{\mathbf{q}\cdot\mathbf{v}}{T}+  \frac{q^2}{2{m}_{\chi}T}+\frac{(\mathbf{q}\cdot\mathbf{v})(\mathbf{q}\cdot\mathbf{v})}{2T^2}\right) 
 \end{equation}
we need to compute:
\begin{eqnarray}
 \frac{{\hat{\mathbf{C}}}_{{\chi}^{0},\rm is}[{f}_{{\chi}^{0}}]}{E} &=& \sum_{(a,b)} \frac{{\tilde{g}}_{Wab}g_{{\chi}^{\pm}}}{256 \pi^5 EE^{\prime}}
\int \frac{{d}^{3}k}{k} f(k) \int {\frac{{d}^{3}{k}^{\prime}}{{k}^{\prime}}} |\bar{M}|^{2}_{ab}
\, \delta \left(E^{\prime}+k^{\prime}-E-k\right) \nonumber\\
&&  \left[\left(f_{{\chi}^{\pm}}(p)e^{\frac{{\Delta m}_{\chi}}{T}}-f_{{\chi}^{0}}(p)\right)- 
\left(\frac{{\Delta m}_{\chi}v^2}{2T}f_{{\chi}^{\pm}}(p)+\frac{\mathbf{q}\cdot\mathbf{v}}{v}\frac{df_{{\chi}^{\pm}}}{dp}+
\frac{\mathbf{q}\cdot\mathbf{v}}{T}f_{{\chi}^{\pm}}(p)\right) e^{\frac{{\Delta m}_{\chi}}{T}} \right.\nonumber\\
&&  +\left(\frac{q^2}{2{m}_{\chi}T}f_{{\chi}^{\pm}}(p)+\frac{(\mathbf{q}\cdot\mathbf{v})(\mathbf{q}\cdot\mathbf{v})}{2T^2}f_{{\chi}^{\pm}}(p)+
\frac{(\mathbf{q}\cdot\mathbf{v})(\mathbf{q}\cdot\mathbf{v})}{vT}\frac{df_{{\chi}^{\pm}}}{dp}+\frac{(\mathbf{q}\cdot\mathbf{v})(\mathbf{q}\cdot\mathbf{v})}{2v^2}{\Delta}_
{\mathbf{p}}f_{{\chi}^{\pm}}\right)e^{\frac{{\Delta m}_{\chi}}{T}}\nonumber\\
&& \left. +\frac{1}{2}\left(\frac{q^2}{v}-\frac{3(\mathbf{q}\cdot\mathbf{v})(\mathbf{q}\cdot\mathbf{v})}{v^3}\right)\frac{df_{{\chi}^{\pm}}}{dp}e^{\frac{{\Delta m}_{\chi}}{T}}\right]
\end{eqnarray}
which is the analogous to the expression for the expansion of the  collisional operator obtained in \cite{Bringmann:2006mu}. As an example we sketch
the calculus of the first term in the square bracket. The invariant amplitude takes the form: 
\begin{equation}
{|\bar{M}|}^{2}_{ab} = {64{G}_{\rm F}^{2}} \left((PK)({P}^{\prime}{K}^{\prime})+(P{K}^{\prime})({P}^{\prime}K)-
({m}_{{\chi}}+\Delta m_{\chi}) {m}_{{\chi}} K{K}^{\prime}\right)
\end{equation}
At the leading order in $T/m_{\chi}$ and ${\Delta m}_{\chi}/m_{\chi}$ we can write:
\begin{eqnarray}
&& \int \frac{{d}^{3}k}{k} f(k) \int {\frac{{d}^{3}{k}^{\prime}}{{k}^{\prime}}} |\bar{M}|^{2}_{ab}
\, \delta \left(E^{\prime}+k^{\prime}-E-k\right) = \int \frac{{d}^{3}k}{k}f(k)\int \frac{{d}^{3}{k}^{\prime}}{{k}^{\prime}}
\delta(\omega +\sqrt{1-{v}^{2}}\Delta m_{\chi}-\mathbf{v} \cdot\mathbf{q}) \cdot \nonumber\\
&& \;\;\cdot \left[\left(2(k-\bold{v} \cdot \bold{k})(k^{\prime}-\bold{v} \cdot \bold{{k}}^{\prime})+\frac{1}{2}({\omega}^{2}-{q}^{2})(1-{v}^{2})\right) 
+\sqrt{1-v^2}\left(-{k}^{\prime} \frac{\mathbf{k} \cdot\mathbf{q}}{{m}_{{\chi}}}-k\frac{{\mathbf{k}}^{\prime} \cdot\mathbf{q}}{{m}_{{\chi}}}+
2\mathbf{v} \cdot\mathbf{k} \,\mathbf{v} \cdot {\mathbf{k}}^{\prime}\frac{\mathbf{v} \cdot\mathbf{q}}{{m}_{{\chi}}} + \right. \right. \nonumber \\
&&\;\;\;\;\;\;\;\;\left. \left. -k\mathbf{v} \cdot {\mathbf{k}}^{\prime}\frac{\Delta m_{\chi}}{{m}_{{\chi}}}-{k}^{\prime}\mathbf{v} \cdot\mathbf{k} \frac{\Delta m_{\chi}}{{m}_{{\chi}}}\right) +\frac{1}{2}\left({\omega}^{2}-{q}^{2}\right)\left(-{v}^{2}\frac{\Delta m_{\chi}}{{m}_{{\chi}}}+(1-\frac{3}{2}{v}^{2})\frac{(\mathbf{v} \cdot\mathbf{q})} {{m}_{{\chi}}}\right) \right] 
\label{eq:a8}
\end{eqnarray} 
The first step of the integration is an average over the directions of $v$. It can be done by use of the following results:
\begin{eqnarray}
 &&\int \frac{d\Omega}{4\pi}\delta(W-\mathbf{v} \cdot\mathbf{q}) =\frac{1}{2vq}\theta({v}^{2}{q}^{2}-{W}^2)   \\
 &&\int \frac{d\Omega}{4\pi}\delta(W-\mathbf{v} \cdot\mathbf{q}){v}^{i} = \frac{1}{2vq}\theta({v}^{2}{q}^{2}-{W}^2)\frac{W}{q^2}{q}^{i}  \nonumber \\
 &&\int \frac{d\Omega}{4\pi}\delta(W-\mathbf{v} \cdot\mathbf{q}){v}^{i}{v}^{j} =
 \frac{1}{2vq}\theta({v}^{2}{q}^{2}-{W}^2)\left(\frac{{W}^{2}-{v}^{2}{q}^{2}}{2{q}^{2}}{\delta}^{ij}+
 \frac{3{W}^{2}-{v}^2{q}^{2}}{2{q}^{4}}{q}^{i}q^{j}\right)  \nonumber \\
 &&\int \frac{d\Omega}{4\pi}\delta(W-\mathbf{v} \cdot\mathbf{q}){v}^{i}{v}^{j}{v}^{k} =
 \frac{1}{2vq}\theta({v}^{2}{q}^{2}-{W}^2) \left(\frac{{v}^{2}{q}^{2}W-W^3}{2{q}^{4}}\left(q^i{\delta}^{jk}+q^j {\delta}^{ik}+q^k{\delta}^{ij}\right)+
 \frac{5{W}^{3}-3{v}^2{q}^{2}W}{2{q}^{6}}{q}^{i}{q}^{j}q^k\right) \nonumber
\end{eqnarray}
where we defined $W \equiv \omega+\sqrt{1-{v}^{2}}\Delta m_{\chi}$. The further 3 integrals in Eq.~(\ref{eq:a8}) are most easily performed in  
in the variables $\omega$, $q$ and $k$. Actually the last step, the integral in the variable $k$ needs to be performed numerically; an analitic 
expression, which traces rather accurately the numerical result, can be obtained by replacing the Fermi-Dirac distribution $f(k)$ with the 
exponential scaling $\exp\left(-k/T\right)$. At the first order in $T/m_{\chi}$ and $\Delta m_\chi/m_\chi$, also remembering that, under our assumptions,
 $p^2/m_{\chi}^2 \simeq T/m_\chi$ this gives:
\begin{eqnarray}
&& \int \frac{{d}^{3}k}{k} f(k) \int {\frac{{d}^{3}{k}^{\prime}}{{k}^{\prime}}} |\bar{M}|^{2}_{ab}
\, \delta \left(E^{\prime}+k^{\prime}-E-k\right) = 
\left[ 4 T^3\left(\Delta m_{\chi}^2+6 \Delta m_{\chi} T+12 T^2\right)\left(1+\frac{{\Delta m}_{\chi}}{m_{\chi}}\right) + \right. \nonumber\\
&& \left. \;\;-2{\Delta m}_{\chi}^{2}T^2\left({\Delta m}_{\chi}+2T\right)\frac{p}{m_\chi}
+\frac{2}{3}T\left({\Delta m}_{\chi}^{4}+3{\Delta m}_{\chi}^{3}T+32{\Delta m}_{\chi}^{2}T^2+114{\Delta m}_{\chi}T^3+144T^4\right)\frac{p^2}{m_{\chi}^{2}}
\right] e^{\frac{{\Delta m}_{\chi}}{T}}\,.
\end{eqnarray}
After computing all integrals, the inelastic scattering contributions to the neutralino and chargino collision terms are then found to be:
\begin{align}
\label{eq:cnscatter}
& \frac{\hat{\mathbf{C}}_{{\chi}_{0},\rm is}\left[f_{{\chi}^{0}},{f}_{{\chi}^{\pm}}\right]}{E}=\sum \frac{2{G}_{\rm F}^{2}{\tilde{g}}_{Wab}{g}_{{\chi}_{\pm}}}{{\pi}^{3}}
\left\{ \left[ 4 T^3\left(\Delta m_{\chi}^2+6 \Delta m_{\chi} T+12 T^2\right)\left(1+\frac{{\Delta m}_{\chi}}{m_{\chi}}\right) \right. \right.
\nonumber\\
& \left. -2{\Delta m}_{\chi}^{2}T^2\left({\Delta m}_{\chi}+2T\right)\frac{p}{{m}_{\chi}}
+\frac{2}{3}T\left({\Delta m}_{\chi}^{4}+3{\Delta m}_{\chi}^{3}T+32{\Delta m}_{\chi}^{2}T^2+114{\Delta m}_{\chi}T^3+144T^4\right)\frac{{p}^2}{m_{\chi}^{2}}\right] \left(f_{{\chi}^{\pm}}-f_{{\chi}^{0}}e^{-\frac{{\Delta m}_{\chi}}{T}}\right)\nonumber\\
&-\frac{8}{3}{\Delta m}_{\chi}T^3 \left({\Delta m}_{\chi}^{2}+6{\Delta m}_{\chi}T+12T^2\right)\left(\frac{{p}^2}{Tm_{\chi}^2}f_{{\chi}_{\pm}}
+\frac{\mathbf{p}\cdot {\nabla}_{\mathbf{p}}f_{{\chi}^{\pm}}}{m_{\chi}}\right)\nonumber\\
& +\frac{2}{3}T^3\left({\Delta m}_{\chi}^{4}+10{\Delta m}_{\chi}^{3}T+60{\Delta m}_{\chi}^{2}T^2+240{\Delta m}_{\chi}T^3+480T^4\right)
\left({\Delta}_{\mathbf{p}}f_{{\chi}^{\pm}}+\frac{\mathbf{p} \cdot {\nabla}_{\mathbf{p}}f_{{\chi}^{\pm}}}{{m}_{\chi}T}+\frac{3}{{m}_{\chi}T}f_{{\chi}^{\pm}}\right)\nonumber\\
& \left. -2T^2 {\Delta m}_{\chi}\left({\Delta m}_{\chi}^{2}+6{\Delta m}_{\chi}T+12T^2\right)\frac{{p}^{2}}{{m}_{\chi}^{2}}f_{{\chi}^{\pm}}\right\}
\end{align}
\begin{align}
\label{eq:ccscatter}
& \frac{\hat{\mathbf{C}}_{{\chi}^{\pm}, \rm is}\left[{f}_{{\chi}^{0}},f_{{\chi}^{\pm}}\right]}{E}=
\sum \frac{2{G}_{\rm F}^{2}{\tilde{g}}_{Wab}{g}_{{\chi}^{0}}}{{\pi}^{3}}
\left\{ \left[  4 T^3\left(\Delta m_{\chi}^2+6 \Delta m_{\chi} T+12 T^2\right)\left(1-\frac{{\Delta m}_{\chi}}{m_{\chi}}\right)
\right. \right. \nonumber\\
& \left. -2{\Delta m}_{\chi}^{2}T^2\left({\Delta m}_{\chi}+2T\right)\frac{p}{{m}_{\chi}}
+\frac{2}{3}T\left({\Delta m}_{\chi}^{4}-4{\Delta m}_{\chi}^{3}T-10{\Delta m}_{\chi}^{2}T^2+30{\Delta m}_{\chi}T^3+144T^4\right)\frac{{p}^2}{m_{\chi}^{2}}
\right] \left(f_{{\chi}^{0}}e^{-\frac{{\Delta m}_{\chi}}{T}}-f_{{\chi}^{\pm}}\right)\nonumber\\
&+\frac{8}{3}{\Delta m}_{\chi}T^3 \left({\Delta m}_{\chi}^{2}+6{\Delta m}_{\chi}T+12T^2\right)\left(\frac{{p}^2}{Tm_{\chi}^2}
f_{{\chi}^{0}}
+\frac{\mathbf{p}\cdot {\nabla}_{\mathbf{p}}f_{{\chi}^{0}}}{m_{\chi}}\right)e^{-\frac{{\Delta m}_{\chi}}{T}}\nonumber\\
& +\frac{2}{3}T^3\left({\Delta m}_{\chi}^{4}+10{\Delta m}_{\chi}^{3}T+60{\Delta m}_{\chi}^{2}T^2+240{\Delta m}_{\chi}T^3+480T^4\right)
\left({\Delta}_{\mathbf{p}}f_{{\chi}^{0}}+\frac{\mathbf{p} \cdot {\nabla}_{\mathbf{p}}f_{{\chi}^{0}}}{{m}_{\chi}T}+\frac{3}{{m}_{\chi}T}f_{{\chi}^{0}}\right)
e^{-\frac{{\Delta m}_{\chi}}{T}}\nonumber\\
&\left. +2T^2 {\Delta m}_{\chi}\left({\Delta m}_{\chi}^{2}+6{\Delta m}_{\chi}T+12T^2\right)\frac{{p}^{2}}{{m}_{\chi}^{2}}f_{{\chi}^{0}}e^{-\frac{{\Delta m}_{\chi}}{T}} \right\}
\end{align}
The ordering of the terms is such that, when integrated over the momentum $\mathbf{p}$ of the neutralino (first equation) or the chargino 
(second equation), the terms on the third and forth row  of the two expressions cancel out 
while, for what regards the others, it can be seen that once summing the two equations one obtains a term proportional to $\left(\frac{p^2}{m_\chi^2}-3\right) 
\left(f_{{\chi}_0}e^{-\frac{\Delta m_\chi}{T}}-f_{{\chi}_{\pm}}\right)$ which cancels out too. The Boltzmann equation in (\ref{eq:singlen}) 
are then obtained after the momentum integration $\mathbf{p}$ and using the fact that 
$T/m_\chi,{\Delta m}_{\chi}/{m_{\chi}} \ll 1$ which allows to keep just the $0^{\rm th}$ order terms in both  (\ref{eq:cnscatter}) and (\ref{eq:ccscatter}).

The equation for the neutralino temperature can be obtained from the second moment of the Boltzmann equation in phase space:
\begin{equation}
 \int  \frac{{d}^{3}p}{{\left(2\pi\right)}^{3}} \, {g}_{\chi^0} {p}^{2} \left(\partial_t-H\mathbf{p}\cdot \nabla_{\mathbf{p}}\right)f_{\chi^0}(p) =
 \int \frac{{d}^{3}p} {{\left(2\pi\right)}^{3}} \,{g}_{\chi^0} \frac{{p}^{2}}{E} \hat{\mathbf{C}}_{{\chi}^{0}} [{f}_{\chi^0}] 
\end{equation} 
The left hand side can be rewritten as:
\begin{equation}
\label{rhs}
 3{n}_{\chi^0}\frac{d{T}_{N}}{dt}+15H{T}_{\chi_0}{n}_{\chi^0}+3{T}_{\chi^0}\frac{d{n}_{\chi^0}}{dt}
\end{equation}
For what regards the right-hand side, we have that the contribution from annihilations, can be computed using the S-wave approximation. In this case 
in fact we can assume that the dipendence of the pair annihilation cross section on the momentum can be neglected and use the same factorization 
implemented when assuming kinetic equilibrium, i.e. :
\begin{equation}
\int \frac{{d}^{3}{p}_{1}}{\left(2\pi\right)^3} \frac{{d}^{3}{p}_{2}} {\left(2\pi\right)^3} g_1 g_2\,(\sigma v) \left({f}_{1}{f}_{2}-{f}_{1}^{eq}{f}_{2}^{eq}\right)
=\langle \sigma v \rangle \left({n}_{1}{n}_{2}-{n}_{1,eq}{n}_{2,eq}\right) 
\end{equation}
with
 \begin{equation}
 \langle \sigma v \rangle= \frac{\int {d}^{3}{p}_{1}{d}^{3}{p}_{2} (\sigma v) {f}_{1}{f}_{2}}{\int {d}^{3}{p}_{1}{d}^{3}{p}_{2} {f}_{1}{f}_{2}} \,. 
 \end{equation}
taking second moments of the distribution functions, one gets a the term:
\begin{equation}
 3 {M}_{{\chi}^{0}}T_{{\chi}_{0}}\left( {\langle \sigma v \rangle}_{{\chi}^{0} {\chi}^{0}} \left({n}_{{\chi}^{0}}^{2}-{n}^{2}_{{\chi}^{0},\rm eq}\right)
+ {\langle \sigma v \rangle}_{{\chi}^{0} {\chi}^{\pm}} \left({n}_{{\chi}^{0}}{n}_{{\chi}^{\pm}}-{n}_{{\chi}^{0},\rm eq}{n}_{{\chi}^{\pm},\rm eq}\right)\right)
\end{equation}
which cancels out against the term on the left hand-side proportional to ${dn_{{\chi}_{0}}}/{dt}$.

\end{document}